%% file: preprint.tex
\documentclass[preprint]{aastex}   
\usepackage{natbib}

\newcommand{\ra}{$R_{\mathrm A}$}
\newcommand{\kms}{km\,s$^{-1}$}
\newcommand{\lxu}{$10^{43}$\,erg\,s$^{-1}$}
\newcommand{\df}{\mbox{DEF}}
\newcommand{\fdef}{$F_\mathrm{DEF}$}
\newcommand{\aas}{A\&AS}       
\newcommand{\nature}{Nature}   

\begin{document}


\title{The HI Content of Spirals. II. Gas Deficiency in Cluster Galaxies}


\author{Jos\'e M. Solanes} \affil{Departament d'Enginyeria
Inform\`atica i Matem\`atiques, Universitat Rovira i Virgili. Carretera
de Salou, s/n; E--43006~Ta\-rra\-go\-na, Spain}
\email{jsolanes@etse.urv.es}

\author{Alberto Manrique} 
\affil{Departament d'Astronomia i
Meteorologia, Universitat de Barcelona. Av.~Diagonal 647;
E--08028~Barcelona, Spain} 
\email{Alberto.Manrique@am.ub.es}

\author{Carlos Garc\'\i a-G\'omez} \affil{Departament d'Enginyeria
Inform\`atica i Matem\`atiques, Universitat Rovira i Virgili. Carretera
de Salou, s/n; E--43006~Ta\-rra\-go\-na, Spain}
\email{cgarcia@etse.urv.es}

\author{Guillermo Gonz\'alez-Casado} \affil{Departament de Matem\`atica
Aplicada II, Universitat Polit\`ecnica de Catalunya. Pau Gargallo 5;
E--08028~Bar\-ce\-lo\-na, Spain} \email{guille@ma2.upc.es}

\and

\author{Riccardo Giovanelli and Martha P. Haynes} \affil{Center for
Radiophysics and Space Research and National Astronomy and Ionosphere
Center\footnote{The National Astronomy and Ionosphere Center is
operated by Cornell University under a cooperative agreement with the
National Science Foundation.},\\ Cornell University; Ithaca, NY 14853}
\email{(riccardo,haynes)@astrosun.tn.cornell.edu}

             
\begin{abstract}
We derive the atomic hydrogen content for a total of 1900 spirals in
the fields of eighteen nearby clusters. By comparing the HI deficiency
distributions of the galaxies inside of and outside of one Abell radius
(\ra) of each cluster, we find that two thirds of the clusters in our
sample show a dearth of neutral gas in their interiors. Possible
connections between the gaseous deficiency and the characteristics of
both the underlying galaxies and their environment are investigated in
order to gain insight into the mechanisms responsible for HI depletion.
While we do not find a statistically significant variation of the
fraction of HI-deficient spirals in a cluster with its global
properties, a number of correlations emerge that argue in favor of the
interplay between spiral disks and their environment. In the clusters
in which neutral gas deficiency is pronounced, we see clear indications
that the degree of HI depletion is related to the morphology of the
galaxies and not to their optical size: early-type and, probably, dwarf
spirals are more easily emptied of gas than the intermediate Sbc--Sc
types. Gas contents below one tenth, and even one hundredth, of the
expectation value have been measured, implying that gas removal is very
efficient. The radial extent of the region with significant gas
ablation can reach up to 2\ra.  Within this zone, the proportion of
gas-poor spirals increases continuously towards the cluster center. The
wealth of 21-cm data collected for the Virgo region has made it
possible to study the 2D pattern of HI deficiency in that cluster. The
map of gas deficiency in the Virgo central area points to an scenario
in which gas losses result from the interaction of the disks with the
inner hot intracluster gas around M87. We also find evidence that
gas-poor spirals in HI-deficient clusters move on orbits more radial
than those of the gas-rich objects. The implications of all these
results on models of how galaxies interact with their environment are
reviewed. Hydrodynamic effects appear as the most plausible cause of HI
removal.
\end{abstract}


\section{Introduction}

In nearby clusters, environmental interactions leave their imprint on
the fragile gaseous disks of the spirals. As a result, comparison of
the neutral hydrogen content of cluster objects with respect to their
field counterparts has been frequently used to evaluate the strength of
these perturbations and to identify their physical origin. Thus far,
the most extensive 21-cm-line studies of high galaxian density regions
have been those of \citet*[hereinafter GH85]{GH85}, who examined a
sample of nine nearby clusters, and those reported by
\citet*[hereinafter HG86]{HG86} and \citet*{HHS88}, who investigated
large compilations of Virgo cluster galaxies. The HI data published by
R.~G.\ and M.~P.~H.\ have been analyzed further by \citet{Dre86},
\citet{Mag88}, and \citet{VJ91}. These investigations argued in favor
of an ongoing interaction between spiral disks and their environment,
but the process responsible for the gas depletion could not be
unambiguously identified.

Because of the renewed interest in the interplay between galaxies and
their surroundings and the substantial body of new cluster HI data that
have been accumulated, conditions are ripe for a new evaluation of the
HI content of galaxies in clusters that can provide fresh input to the
debate about the extent to which the environment influences the
evolution of galaxies. The firsts steps in this direction were taken a
few years ago in a paper by \citet*[hereinafter Paper~I]{SGH96}, which
was focused on the derivation of Malmquist-bias free estimates of the
HI mass standards for the different morphological subgroups of luminous
spirals. In the present paper, we exploit a large dataset of HI
observations of spiral galaxies in the regions of eighteen nearby
clusters with the aim of unmasking the cause of the observed gas
deficiency.

The outline of the paper is as follows. The next section describes the
criteria adopted for the selection of the cluster galaxy samples and
the manner in which the HI deficiency is measured. The Virgo cluster
and a composite formed by the remaining HI-deficient systems are used
in \S~\ref{pattern} to investigate the spatial pattern of HI
deficiency. The dependence of HI deficiency both on the main global
properties of clusters, and on the morphology and size of the galaxies
is explored in \S~\ref{clusprop} and \ref{galprop},
respectively. Finally, \S~\ref{orbits} examines the orbits of the
spirals in HI-deficient clusters according to their gaseous
content. The implications of the results of such analyses are discussed
in \S~\ref{discussion}. The paper concludes with a summary of our
findings. For all calculations involving the Hubble parameter we take
$H_0=100\,h$ km\,s$^{-1}$\,Mpc$^{-1}$.

\section{The Cluster Sample}\label{sample}

The galaxies used in the present study have been extracted from the
all-sky database of nearby galaxies maintained by R.~G.\ and M.~P.~H.\
known as the Arecibo General Catalog (AGC). Apart from several entries
listing 21-cm line information, the AGC contains an extensive
compilation of galaxy parameters from optical observations, such as
morphologies and apparent sizes. A large number of galaxies have
morphological types and visual estimates of their angular diameters
listed in the {\it Uppsala General Catalog of Galaxies\/}
\citep[UGC,][]{Nil73}. For non-UGC objects, these properties have been
obtained from the visual examination of the Palomar Observatory Sky
Survey prints. The typical uncertainty in the morphological
classification is $+/-$ one Hubble type, although for some individual
galaxies, especially the ones with the smallest angular sizes, the
errors may be larger. Apparent diameters are estimated in bins of
$0\farcm1$ for dimensions larger than $1\arcmin$ and in bins of
$0\farcm05$ for dimensions below that value. Because the eye measures
at about the level of a \emph{face-on} isophotal radius, independently
of inclination \citep{Gio94}, visual diameters have not been corrected
for the inclination of the parent galaxy (internal absorption). We also
neglect the effects of galactic extinction on apparent dimensions. The
AGC also contains information on heliocentric radial velocities from
radio and/or optical wavelengths (the radio measurement taking always
preference over the optical one when the two are available).

Following our previous work in this subject, HI deficiencies for
individual galaxies have been quantified by means of a parameter $\df$
that compares, in logarithmic units, the observed HI mass, $h^2M_{\rm
HI}^{\rm obs}$, inferred from the corrected HI flux, with the value
expected from an isolated (i.e.\ free from external influences) galaxy
of the same morphological type, $T^{\rm obs}$, and optical linear
diameter, $hD_{\rm opt}^{\rm obs}$, calculated from the major visual
dimension (for details, see \citealt*{HG84}, hereinafter
HG84). Specifically:
\begin{equation} \label{def2} 
\df=\;\langle\log M_{\rm HI}(T^{\rm obs},D_{\rm opt}^{\rm
obs})\rangle-\log M_{\rm HI}^{\rm obs}\;,
\end{equation}  
so positive values of $\df$ indicate HI deficiency. In eq.~(\ref{def2})
the HI mass is expressed in solar units and the optical diameter in
kpc. For the expectation value of the (logarithm of the) HI mass, we
use the maximum likelihood linear regressions of $\log(h^2M_{\rm HI})$
on $\log(hD_{\rm opt})$ inferred from the field galaxy sample
summarized in Table~2 of \citeauthor{SGH96}. Because the standards of
normalcy for the HI content are well defined only for the giant spiral
population (Sa--Sc), we have excluded from the present study earlier
Hubble types, as well as all galaxies unclassified or with peculiar or
very disturbed morphologies. Nonetheless, HI mass contents for the few
HI-rich Sd's, Sdm's, and Magellanic-type irregulars included in our
samples have been calculated from the relationship inferred for the
Sc's following the results of \citeauthor{HG84}. Because the AGC data
have been gathered from a wide variety of sources, data quality is
inhomogeneous. While we have been careful to select only those galaxies
for which a meaningful HI measure can be calculated, the reader should
be aware that the applicability of the assembled dataset is essentially
statistical.  Imprecisions in the diameter measures and morphological
type assignments, together with the built-in distance dependence of the
$M_{\rm HI}-D_{\rm opt}$ relationship, make determinations of the HI
deficiency for individual objects uncertain.

To be assigned to a given cluster field, a galaxy must lie within a
projected distance of 5\ra, i.e.\ within 7.5\,$h^{-1}$ Mpc, of the
cluster center and have a radial velocity which is separated from the
recessional velocity of the cluster no more than $\sim 3$ times its
average velocity dispersion. Since we are especially interested in the
central portions of clusters where environmental influences are
strongest, we have only selected those clusters with at least 10
galaxies (of types Sa--Sdm/Irr) with good HI detections located within
the innermost 1\ra\ (1.5\,$h^{-1}$ Mpc) radius circle.

Of all the cluster fields sampled in the AGC a total of eighteen
satisfy all the above constraints. These are the sky regions centered
on the ACO \citep*{ACO89} clusters A262, A397, A400, A426 (Perseus),
A539, A779, A1060 (Hydra I), A1367, A1656 (Coma), A2063, A2147, A2151
(Hercules), and A3526 (Centaurus30), on the clusters of Virgo, Pegasus,
Cancer, and Pisces, and on the group of galaxies around NGC507,
hereinafter referred to as N507. Table~1 lists, for the above galaxy
concentrations, the following quantities: cluster name in column~(1);
in columns~(2) and (3) the cluster center coordinates referred to the
1950 epoch (mostly taken from the \emph{Einstein} catalog of
\citealt{JF99}; when no X-ray observations are available---as for
N507---we use the peak of the galaxy distribution); our velocity filter
used to define cluster membership in column~(4); in column (5) the
Abell radius of the cluster expressed in degrees, inferred from its
cosmological distance; and in columns (6) and (7), respectively, the
total number of S's meeting the membership criteria within 1 and 5\ra,
except for the Virgo cluster region for which a maximum radial cutoff
of 3\ra\ has been imposed to avoid dealing with large angular
separations. The sky distributions of the galaxies belonging to each
one of the eighteen cluster regions are plotted in Figure~1. Figure~2
shows the histograms of the distribution of the measured values of
$\df$ according to equation~(\ref{def2}) for these same regions. In the
histograms, the filled areas illustrate the distribution for the
galaxies within 1\ra---for which we adopt the cluster distance---,
while the unfilled areas correspond to those objects at larger
radii. Apart from some expected contamination by outliers, it is
evident from this latter figure that, while the outer distributions
tend to be bell-shaped, exhibit a dispersion comparable to the value of
0.24 measured for isolated galaxies (see \citeauthor{SGH96}), and peak
around zero $\df$, the central galaxies of the majority of the clusters
show evidence for strong HI deficiency. The most notable case is that
of Virgo, for which some of the inner galaxies have HI masses up to two
orders of magnitude below their expectation values. Note that some of
the datasets include a few galaxies undetected in HI but for which a
reliable upper limit of their HI content has been calculated (see again
\citeauthor{HG84} for further details). In the calculations of $\df$
made through this paper, non-detections always contribute with their
nominal lower limit of deficiency.

We choose to define a cluster as HI deficient when a two-sample
Kolmogorov-Smirnov (KS) test gives a probability of less than 10\% that
its observed inner and outer distributions of $\df$ are drawn from the
same parent population. Although it could be argued that the small size
of the inner distributions makes some comparisons uncertain, the
adopted definition has the advantage of being fully objective. Indeed,
the results of the KS test confirm essentially the visual impression:
the central spiral populations of twelve of the cluster fields in our
dataset, Pegasus, Virgo, A262, A397, A400, A426, A779, A1060, A1367,
A1656, A2063, and A2147 (identified in Fig.~2 by an asterisk after the
name) have statistically significant reduced HI contents. Among these,
only for A400 and A1060---two clusters with rather complex central
velocity distributions \citep[see e.g.][]{Bee92,FM88}--- is it not
possible to reject the null hypothesis at better than the $5\%$ level
of significance. For the other six galaxy concentrations, Cancer, N507,
Pisces, A539, A2151, and A3526, the presence of objects with large HI
deficiencies in the central regions is an exception of the norm.  The
results of the KS test are listed in the last column of Table~1.

From Figure~2 and Table~1, it is also readily apparent that, in spite
of the stricter radial cutoff applied, the galaxy sample of the Virgo
region is, because of its proximity, the largest. The difference in
size is especially striking when one looks at the central 1\ra-radius
circle: within this zone, clusters other than Virgo contain on the
average 20 objects, while this latter system has nearly 11 times more
galaxies. Thanks to this substantial wealth of HI data, Virgo is the
only single cluster of our dataset for which it is possible to
investigate in detail the relationship of HI deficiency with the
position, morphology and kinematics of the galaxies (see
\S\S~\ref{vir3w}, \ref{galprop}, and \ref{orbits}). The remaining
galaxy samples are useful to investigate the HI deficiency only in an
overall sense, while for more detailed analysis, one must resort to the
combination of the individual datasets to increase the statistical
reliability of the results.

\section{The Spatial Pattern of HI Deficiency}\label{pattern}
 
\subsection{The Virgo Cluster}\label{vir3w}

The Virgo cluster is the nearest large-scale galaxy concentration which
offers the possibility of exploring the manifestations of environmental
effects on galaxies with greatest detail. Nowadays, thanks to the large
amount of 21-cm data accumulated, the case supporting the HI deficiency
of the spirals in the core of this cluster is solidly established
(e.g.\ \citealt*{DL73,vdB76,Gio83,GH83,CBF86}; \citeauthor{HG86};
\citealt*{Gui87}). It has also been clearly demonstrated that the sizes
of the gaseous disks of the HI-poor Virgo spirals are reduced with
respect to their field counterparts
\citep*[e.g.][]{Hel81,Gio83,GH83,War88a,War88b} in an amount that
increases with decreasing distance to the cluster center---marked by
the giant elliptical galaxy M87 \citep{War86,Cay94}. The selective
sweeping of HI in the outer portions of the disks points to gas removal
mechanisms initiated by the surrounding intergalactic medium (IGM).

Figure~3 shows the 2D adaptive map of the gas deficiency in the central
cluster region. We have restricted our original Virgo sample to the
subset of 187 galaxies with good 21-cm measures located in a region of
size $10\degr\times 10\degr$ bounded by $12^{\mathrm h} 7^{\mathrm
m}\le \mbox{R.A.}\le 12^{\mathrm h} 47^{\mathrm m}$, $6\fdg 5\le
\mbox{Dec.}\le 16\fdg 5$. This region is centered on (and covers most
of) the classical Virgo I Cluster area \citep{dVdV73}. When
constructing the HI-deficiency map, we have also taken into account the
close proximity and dynamical complexity of the Virgo cluster. The lack
of correspondence between the observed radial velocity and distance in
regions detached from the Hubble flow is a source of scatter that, for
the nearby Virgo system, largely dominates the calculation of intrinsic
parameters that have a built-in distance dependence, such as our
deficiency estimator (eq.~[\ref{def2}]): $M_{\rm HI}\propto D_{\rm
opt}^{n}$, with $n\approx 1.7$ for Sc's and $n\approx 1.2$ for earlier
spiral types (\citeauthor{SGH96}). To estimate the contribution to the
HI-deficiency map of spurious fluctuations caused by possible erroneous
distance assignments, we have derived a second map by using a
distance-independent approximation to equation~(\ref{def2}) based on
the difference between the expected and observed logarithm of the mean
HI surface density, $\overline{\Sigma}_{\rm HI}$, that is:
\begin{equation} \label{def1} 
\df=\langle\log \overline{\Sigma}_{\rm HI} (T^{\rm obs})\rangle-\log
\overline{\Sigma}_{\rm HI}^{\rm obs}\;,
\end{equation}
with $\overline{\Sigma}_{\rm HI}=F_{\rm HI}/a_{\rm opt}^2$, and where
$F_{\rm HI}$ represents the corrected HI flux density integrated over
the profile width in units of Jy\,\kms\ and $a_{\rm opt}$ the apparent
optical diameter in arcmin (see also eqs.~[3], [6], and [7] of
\citeauthor{SGH96}). Note that $\overline{\Sigma}_{\rm HI}$ is a hybrid
quantity since it uses the optical disk area. Figure~3 shows that,
except for a global mild enhancement of $\sim 0.15$ units in the values
of the HI deficiency parameter most likely caused by background
contamination, the original radio map (Fig.~3a) exhibits a spatial
distribution of deficiency almost identical to that of its
distance-independent approximation (Fig.~3b). This result allow us to
conclude that all the fluctuations depicted in the Figure~3a reflect
true local variations in the gas content of the galaxies.

Several structures emerge clearly from the cluster HI deficiency
distribution.  The zone with the maximum gas deficiency coincides with
both the peak of X-ray emission and the main density enhancement known
as Cluster A \citep*{BPT93}. This is a double system comprising the
subclusters centered on the giant ellipticals M87 and M86, which seem
to be in the process of merging \citep*{SBB99}. Five other distinct
gas-deficient patches appear to be radially connected with the central
one. Two of them are located along the N-S direction: to the South, the
HI deficiency extends towards the clump dominated by M49 (Cluster B in
\citealt{BPT93}); to the North, there is a mild increase of gas
deficiency around the spiral M100. Along the EW axis, the distribution
of HI deficiency is dominated by a region of strong gas depletion to
the East. This EW asymmetry in the HI content has also been observed at
X-ray wavelengths by \citet{Boh94}, who found that the faint Virgo
X-ray emission can be traced out to a distance of $\lesssim 5\degr$,
except in the western side where the emission falls off more
steeply. On the other hand, the position of the eastern local maximum
of deficiency is located about one and a half degrees South of the peak
of the density enhancement known as Cluster C around the pair of
galaxies M59 and M60. The last two zones of important HI depletion are
found near the periphery of the surveyed region, where no X-ray gas is
detected, in the areas of the background galaxy concentrations known as
the M cloud in the NW, and the W' group and (northernmost part of the)
W cloud in the SW.

We comment upon the implications of the maps depicted in Figure~3 on
the possible origin of gas deficiency in this cluster at the end of 
\S~\ref{discussion}.

\subsection{The Radial Variation of HI Deficiency}\label{radial}

A well-known property of the HI deficiency pattern in clusters is its
radial nature. Previous studies of cluster galaxy samples have already
revealed that gas-poor objects are more abundant in the centers of
clusters than in their periphery (\citealt*{GCH81,Sul81,BSS84};
\citeauthor{GH85}; \citeauthor{HG86}; \citealt*{Mag88,Bra00}). With our
new data, this effect can also be observed in the greyscale maps
depicting the HI deficiency distribution in the dynamically unrelaxed
Virgo core (Fig.~3). In spite of the irregular distribution of the
galaxies, the shade intensity of these maps, which is proportional to
the deficiency measure, grows toward the position of M87, where the
density of the environment is highest.

A more precise characterization of the radial behavior of HI deficiency
is obtained by combining into a single dataset the HI measures for
spirals in all the clusters which show HI deficiency other than Virgo
(see \S~\ref{sample}), with their clustercentric distances normalized
to \ra. This composite sample of eleven HI-deficient clusters allows us
to trace deficiency out to projected distances from the cluster center
of 5\ra\ in much greater detail than the (relatively) small samples of
the individual clusters, while at the same time reduces possible
distortions caused by substructure and asphericity. Virgo is excluded
from this composite cluster because of its smaller radial extent and
the fact that its much larger dataset would dominate the composite
cluster. Because the established upper limits of gas content for
undetected galaxies in nearby clusters are more stringent than in
distant ones, it is not appropriate to use average values of $\df$ to
characterize gas removal. We adopt instead a measure based on the
relative populations of deficient and normal spirals, much less
sensitive to the presence of censored data. In Figure~4, we show the
variation in the fraction of spirals with $\df>0.30$ per bin of
projected radial distance, in Abell radius units, for our composite
HI-deficient cluster. The radial dependence of HI deficiency is clearly
evident for $r\lesssim 2$\ra: the percentage of gas-poor spirals
increases monotonically up to the center. Beyond this projected
distance, however, the fraction of gas-deficient disks remains constant
around a value of $\sim$10--15\%, a value consistent with the fraction
of field spirals with $\df>0.30$ expected from a Gaussian distribution
of values of this parameter with an average dispersion of $0.24$
(\citeauthor{SGH96}). Similar results are obtained when the deficiency
threshold is increased to 0.48 (equivalent to a factor of three
decrease in $M_{\rm HI}$). We have also included in Figure~4 a second
panel showing the variation of HI deficiency with projected radius. It
provides visual verification of the fact that, at large clustercentric
distances where gas-deficient galaxies are scarce and the contribution
of non-detections negligible, the distributions of HI content at
different radii are in excellent agreement both in terms of location
and scale with that of field galaxies. This latter result supports
further the statistical reliability of our measures of $\df$ through
eq.~(\ref{def2}).

The same two previous plots are repeated in Figure~5 for the
superposition of the clusters which are not deficient in HI. It can be
seen that the spirals belonging to these systems show HI contents which
are both essentially independent of their clustercentric distance and
typical of the field population. Comparison of the results in Figures~4
and 5 reinforce also our cluster subdivision into HI-deficient and
``HI-normal'' systems and therefore the soundness of the procedure
followed for such classification.

It is clear from Figure~4 that the influence of the cluster environment
on the neutral gas content of galaxies can extend beyond one Abell
radius (plots of the radial variation of HI deficiency for individual
clusters, not shown here, indicate that the extent of the zone of
significant HI deficiency fluctuates significantly from cluster to
cluster; see also \citeauthor{GH85}). On the other hand, \citet{Bal98},
using [OII] equivalent width data, have found that the mean star
formation rate in cluster galaxies, another property sensitive to
environmental effects, shows signs of depression with respect to the
field values at distances around twice the $R_{200}$ ``virial''
radius\footnote{In a typical nearby rich cluster the values of
$R_{200}$ and the Abell radius are similar: simple calculations show
that these two scales coincide for a $z=0$ cluster with a velocity
dispersion of 866~\kms\ \citep*{CYE97}}. These results seem to pose a
problem for stripping mechanisms which require high environmental
densities to be effective. The reduction of both the gas content and
the star formation rate of these outer cluster galaxies can be
achieved, however, if they are on strongly eccentric orbits that have
carried them through the cluster center at least once (see
\S\S~\ref{orbits} and \ref{discussion}). From the theoretical point of
view, this possibility is supported by the simulations of hierarchical
structure formation by \citet{Ghi98} and \citet{RdS98}, and by recent
direct models of the origin of clustercentric gradients in the star
formation rates within cold dark matter cosmogonies by \citet*{BNM00}.

\section{HI Deficiency and Cluster Properties}\label{clusprop}

The compilation of the principal properties of our clusters from the
literature presents a number of difficulties, most important of which
is the heterogeneity of the available data. Thus, in spite of the fact
that for some of the clusters it is possible to draw their relevant
parameters from detailed individual studies, data were extracted mainly
from large cluster catalogs. By adopting this approach, we insure that
the information available for each cluster property is homogeneous. The
global parameters available for most of the systems in our sample are
summarized in Table~2. These are: three different estimates of the
X-ray luminosity in the 0.01--80~keV (bolometric), 0.5--3~keV (soft),
and 2--10~keV (hard) bandpasses, in columns (2)--(4), respectively;
spectral X-ray temperature (col.~5); radial velocity dispersion
(col.~6); Abell galaxy number counts (col.~7); and spiral fraction
(col.~8). Table~2 is completed with a global measure of the degree of
gas removal in each cluster computed as the ratio of the number of
spirals with $\df>0.30$ found within 1\ra\ of the cluster center to
that of all objects of this type observed in HI within the same
region. This ``HI-deficient fraction'', \fdef, and its associated
Poissonian $1\sigma$-error are indicated in column~(9). The reference
sources have been appended to the tabular information.

The fact that the characteristics of our clusters vary widely suggests
that it is worth investigating correlations between the overall degree
of gas depletion and the global cluster properties that reflect the
strength of the environmental perturbations on the gaseous
disks. Figure~6 shows the four X-ray parameters included in Table~2
plotted against \fdef, while in Figure~7 this fraction is compared with
the three optical properties. A fourth panel in this last figure
compares the bolometric X-ray luminosity with the total spiral
fraction. We note that all plots involving the parameter \fdef\
resemble essentially scatter diagrams with no significant
correlations. We have tested different thresholds of HI deficiency
without finding appreciable changes. Neither does the use of the
distance-independent approximation to $\df$ given by eq.~(\ref{def1})
modify the results significantly, implying that the possible
contamination by interlopers has a negligible contribution to the
scatter of the relationships. Only for the Virgo cluster does the value
of \fdef\ drop significantly (from 63 to 46\%), indicating that, for
this system, most of the interlopers are located in the background.

We recall at this point that in the original investigation of
\citeauthor{GH85}, suggestive though inconclusive indications of a
trend toward greater HI-deficient fraction among clusters with high
X-ray luminosity (in the 0.5--3.0~keV band) were found. This
relationship, however, is not corroborated with the present larger
dataset. One reasonable explanation for the lack of any discernible
correlation is the possible transmutation of some of the swept spirals
into lenticulars, thereby weakening the relationships we are
investigating by reducing the fraction of HI deficient galaxies to a
greater degree for the strong X-ray clusters than for the weak
ones. This possibility is suggested by the very strong anticorrelation
between the total spiral fraction and the X-ray luminosity ($r\le
-0.90$ for all three wavebands; see also \citealt*{Bah77} and
\citealt*{ES91}) in the bottom right panel of Figure~7, implying that
the fraction of lenticulars is correlated with X-ray luminosity.

One caveat that should be mentioned here is that possible selection
effects and the incompleteness of some of our cluster galaxy
samples---not included in the formal random error bars but which may
blur significantly the results---might also explain the lack of good
correlations between the fraction of HI-deficient spirals in clusters
and the global properties of those systems. Note, for instance, that
X-ray luminous clusters are more susceptible to incompleteness effects
since they have a lower fraction of spirals. In an attempt to reduce
the scatter of the plots, we have excluded from the analysis those
samples containing fewer than than 20 objects (identified in Figures~6
and 7 with light gray circles), which are the most likely affected by
problems related to the small sample size. Systems in this restricted
dataset certainly show signs of a possible relationship between \fdef\
and the cluster X-ray luminosity in the 0.5--3.0~keV range (the linear
correlation coefficient, $r$, is equal to 0.55), but there is no
evidence for this trend in the other two X-ray windows. We argue that
the results of the present exercise are not fully conclusive and
require further investigation by means of still larger and more
complete 21-cm-line investigation of galaxies in cluster fields.

\section{HI Deficiency and Galaxy Properties}\label{galprop} 

Important clues to the nature of HI deficiency can also be inferred
from studying the variation of the degree of gas depletion with the
intrinsic properties of the galaxies. In this section, we investigate
possible connections between the gas content and the morphology and
optical size of the disks.

\subsection{Variation with Morphological Type}\label{type}

The increase of gas deficiency towards early morphological types was
first noted for the Virgo giant spiral population by \citet{Sta83} and
\citet{GR85}. This trend was bolstered by \citet{CBF86}, who found that
the HI deficiency in the Virgo cluster spirals increases monotonically
along the Hubble sequence from Sc to the earliest spiral
types. \citeauthor{GH85}'s HI data were also found to obey a gas
deficiency-morphology relationship by \citet{Dre86}, which was
independent of the projected radial distance of the galaxies from the
cluster center.

With our new, extensive data, the variation of HI deficiency with
morphological type can now be reviewed in a far greater detail and put
on a much firmer ground. We restrict the analysis to the twelve
HI-deficient clusters to emphasize the significance of the results. As
before, the data are grouped in a single composite cluster, except for
the Virgo galaxies which are treated separately. We begin by
presenting, in Figure~8, the bar charts of the percentage of galaxies
inside 1\ra\ at a given morphology with deficiency parameter $\df>0.30$
in the Virgo and the composite samples. Hubble types have been replaced
by its numerical $T$-code by \citeauthor{HG84}, which for the galaxies
in the present study runs from $T=3$ for Sa's to $T=9$ for Sd--Sm and
irregular galaxies. No distinction has been made between normal and
barred spirals. Comparison of the two bar charts shows, in the first
place, that the Virgo cluster exhibits a notably larger fraction of
gas-deficient galaxies for any given morphological class, a result
which is simply due to the fact that we are sampling farther down the
HI mass function in Virgo than in the more distant
clusters. Differences in the normalization aside, the bar charts
confirm that for a spiral, the likelihood of being HI deficient depends
on its morphology. Both the Virgo and the composite cluster sample
share a common pattern: a roughly gradual descent of the fraction of
HI-deficient galaxies as the Hubble type goes from Sa to Sc, by a total
amount of $\sim 40\%$, which levels off for the latest types. The only
discrepancy arises in the very latest morphology bin, which shows a
noticeable recovery of the deficiency fraction for Virgo and a sharp
drop for the composite cluster. In this case, we assign more
credibility to the Virgo data since very gas-poor dwarf galaxies are
underrepresented in the more distant clusters.

We have also produced for the above two cluster galaxy subsets the
distributions of values of the parameter $\df$ separately by
morphological type. Given the differences in the morphological
composition between Virgo and the composite cluster (indicated by the
numbers inside the bars in Fig.~8), we find it preferable to adopt a
distinct type grouping for the two samples, so that we give priority to
the reduction of statistical noise. A major feature of the plots
displayed in Figure~9 is the strong positive skewness of all
distributions. In the Virgo cluster, the Sa's exhibit the most radical
behavior: 16 out of the 21 galaxies of this type have HI deficiencies
larger than 1, i.e.\ a factor of 10 decrement over the typical HI
mass. Indeed, all Virgo galaxies with $\df\ge 2$ belong to this
subclass (it should be noted that the high fraction of non-detections
for this type indicates even higher deficiency values than this
limit). Most of the remaining strongly deficient galaxies ($\df\ge 1$)
in the Virgo sample, including the rest of the non-detections but one,
belong to the Sdm/Irr class. This result demonstrates that the dwarf
types, as well as the Sa's, are not only more likely to be deficient in
HI than the intermediate spirals, but they also have a higher gas
deficiency. The distributions for the composite cluster, on the other
hand, show a progressive increase in the positive skewness and boxiness
towards the early types, but the differences among the histograms are
less pronounced than for the Virgo data. There are also very few
galaxies with $\df\ge 1$. The fractions of non-detections, however, are
seen to increase towards the two extremes of the morphological
range. No doubt selection biases against galaxies with very low HI
masses are responsible for the abrupt cutoff of the high-HI-deficiency
tails of the distributions. Clearly, the Virgo sample is much deeper
and complete than any of the other cluster galaxy samples under
scrutiny.

Yet, the possibility remains that the observed correlation between HI
deficiency and morphology might reflect nothing more than the
well-known morphological segregation of cluster galaxies. In other
words, the larger deficiency of the earlier types might be explained
simply by their more central locations. Figure~10 demonstrates that
this is not the case. In this plot, we reproduce the radial run of the
HI-deficient fraction for the composite HI-deficient cluster sample of
Figure~4 but separating the early ($T$:\,$3\div 6$) and late
($T$:\,$7\div 9$) spiral type subsets. From this graph, one sees that,
inside the region of influence of the cluster environment ($r\lesssim
2$\ra), early-type galaxies have systematically higher gas
deficiencies, \emph{at any projected radius}, than the late
types. Identical results, although with more abrupt radial variations
due to the spatial lumpiness of the cluster, are found for the Virgo
sample. We can also rule out strong projection effects, which would
affect preferentially the late types, for two reasons. The first
argument results from the radial velocity filters applied in the
selection of the galaxy samples. The second reason has to do with the
fact that the contribution of a (presumably) uniform distribution of
outliers would be less noticeable in the centermost radial bins, where
the cluster density is the highest. Contrary to these expectations,
Figure~10 shows that the difference between the HI-deficient fractions
of the early- and late-type populations increases gradually towards the
cluster center. We conclude that the observed correlation of HI
deficiency and morphology is not a secondary effect of the spatial
segregation of the galaxies, but reflects the interplay between the
intrinsic characteristics of these objects and the physical mechanism
behind HI depletion.

\subsection{Variation with Galaxy Size}\label{size}

The analysis by \citet{VJ91} of the central galaxies in four of the
HI-deficient clusters identified by \citeauthor{GH85}---A262, A1367,
Coma, and Virgo---revealed an apparent tendency for HI deficiency to
increase with increasing optical galaxy size, a result arising
essentially from the galaxies located at relatively large projected
radial distances from the cluster center ($r>0.75$\ra). According to
those authors, this observational result was difficult to reconcile
with galaxy-intracluster medium interactions, such as ram pressure
stripping and transport processes, even if significant mass segregation
was invoked. Therefore, it becomes quite important to review with our
new data this possible relationship between optical size and gas
deficiency, because its confirmation would put serious strain on some of
the most popular mechanisms of gas depletion.

In order to emphasize the contribution of cluster members, we have
restricted this study to the galaxies within a projected distance of
1\ra\ from the center of the twelve HI-deficient clusters identified in
our catalog. The linear optical sizes of the galaxies are calculated
from their major angular diameters as given in the AGC and from the
mean cluster distances (see \S~\ref{sample}). The optical diameters are
then distributed in logarithmically spaced bins. As in previous
sections, we have investigated separately the variation of the
HI-deficient fraction as a function of galaxy size for the Virgo
dataset and for the composite sample formed by the combination of the
remaining HI-deficient clusters.  The top two panels of Figure~11
depict the corresponding histograms. Contrary to the results reported
in \citet{VJ91}, we find no obvious relationship between HI deficiency
and optical size. A $\chi^2$-test corroborates that, statistically,
there is no significant difference between the observed distributions
and the uniform. We have verified that the results of the analysis are
insensitive to alterations in the binning and to the exact deficiency
criterion adopted.

We know from the results of the preceding sections that HI deficiency
correlates with the morphology of the galaxies and their projected
distance from the cluster center. As a simple method of subtracting the
contributions of these two factors to the correlation that is being
investigated, we have broken down the original histograms by
morphological type (EARLY or LATE) and radial position (INside or
OUTside a circle of $r=0.5$\ra). The results depicted, respectively, in
the middle and bottom panels of Figure~11 correspond only to the Virgo
cluster, but they can be extended to the composite cluster
too. Inspection of these plots shows no noticeable differences between
the behaviors of the histograms of each partition, which are all again
statistically flat, except for the expected overall increase in the HI
deficiency of the EARLY and IN subsets. Notice also that the dynamical
range of the optical diameters for this dataset is much wider than that
of the composite sample, justifying the independent analysis of the
Virgo cluster.

Two arguments can be invoked to explain the different results obtained
by \citet{VJ91}. The most important is the fact that the deficiency
parameter adopted by those authors had a relatively strong residual
dependence on the optical diameter, because it relied upon the
comparison of the mean values of the hybrid surface density of HI (as
in eq.~[\ref{def1}]). As shown in \citeauthor{SGH96},
$\overline{\Sigma}_{\rm HI}$ decreases significantly with increasing
disk size for all the giant spiral types but the latest. Since the
galaxy population in dense environments is biased against late disks,
investigations of the HI deficiency in galaxy clusters based on the
constancy of this quantity are likely to overestimate (underestimate)
the gas deficiency of the largest (smallest) objects. At the time the
earlier study was made, however, the standards of HI content available
predicted that $M_{\rm HI}\propto D_{\rm opt}^{1.8}$ for the entire
spiral population (\citeauthor{HG84}). This prompted \citeauthor{VJ91}
to neglect the intrinsic size dependence of $\overline{\Sigma}_{\rm
HI}$ as insufficient to explain the observed trend. A second factor
that might have contributed to generate the false relationship is the
small size of the galaxy samples, which forced those authors to operate
with a reduced number of intervals dominated by strong numerical
uncertainties.

\section{HI Deficiency and Galaxy Orbits}\label{orbits}

The hypothesis that HI-deficient galaxies lose their interstellar HI at
small distances from cluster cores but can still be found at large
radial distances (\S~\ref{radial}) suggests that the HI-deficient
objects follow highly eccentric orbits. We now investigate the
trajectories of the galaxies in the central regions ($r\le\!1$\ra) of
Virgo and the composite HI-deficient cluster as a function of their gas
contents and morphologies. It should be noted that this approach is
less severely affected by the randomizing effects of geometric
projections than analyses of the distributions of HI deficiency vs.\
projected velocity, which have failed to provide any evidence that
these two quantities are interrelated (\citeauthor{GH85,HG86}).

Information on the eccentricity of galaxy orbits can be extracted from
the radial run of the line-of-sight (los) velocity dispersion. While
the inverse problem of recovering orbital information from radial
velocity data only is undetermined, the direct problem is not. Thus, a
system with galaxies predominantly in radial orbits necessarily
produces an outwardly declining $\sigma_\mathrm{los}$
profile. Accordingly, the observation of such a trend in a cluster
\emph{is consistent} with radial orbits, while the opposite behavior
suggests instead that the galaxy orbits are largely circular. On the
other hand, a roughly constant velocity dispersion with projected
radius is characteristic (although not exclusive) of an isotropic
distribution of velocities.

We have applied a procedure based on the deconvolution method developed
by \citet{SanS89} and \citet{SalS89} to determine the velocity
dispersion curves. This technique---which can be used to infer the
radial profile of any positive quantity in systems with circular or
self-similar symmetry---presents several advantages over the crude
annular binning used by \citet{Dre86} in his investigation of the
orbital parameters of \citeauthor{GH85}'s galaxies. Among its
interesting features are, for instance, its suitability for small
samples since the binning of the data is avoided and the fact that it
yields a quasi-continuos numerical solution (i.e.\ known with an
arbitrarily small sampling interval). We refer those interested in the
fundamentals of this method to the references provided above.

The dependence of the los velocity dispersion on (projected) radius
from the cluster center is obtained simply by taking the square root of
the ratio of the specific kinetic energy---given by the observed
peculiar velocity squared---and the number density profiles of the
galaxies. In order to remove from this kinematic profile the disturbing
effects of subclustering, each system has been first ``circularized''
by performing azimuthal scramblings of the observed galaxy positions
around the cluster center. In this manner, the distribution of
clustercentric distances is preserved, while at the same time any
possible subclumps existing in the original structure are destroyed. In
addition, the observed peculiar velocities of the galaxies are scaled
to the average los velocity dispersion of their parent cluster (see
Table~2). This normalization, which is relevant for the composite
dataset, has been adopted to give equal weight to identical fractional
variations in the velocity dispersion coming from galaxies in different
clusters, as well as to avoid artefacts caused by possible fluctuations
in the degree of completeness of the galaxy samples according to the
clustercentric distance. The mean radial profiles of the normalized los
velocity dispersion, $\sigma^\ast_\mathrm{los}$, for six different
galaxian subpopulations, calculated from 100 circularized realizations
of the Virgo and composite clusters, are displayed in Figure~12. A
low-passband filter with a resolution length of 0.3\ra\ has been
applied to each individual simulation to wash out the noise from
non-significant statistical fluctuations.

We see in the bottom panel of Figure~12, which depicts the curves
corresponding to the composite HI-deficient cluster, that the
normalized velocity dispersion for the spirals with the strongest gas
deficiencies ($\df\ge 0.48$) drops significantly in a manner consistent
with radial orbits \citep[see also][]{Dre86}. The curve for the
gas-rich objects ($\df\le 0$) decreases too with increasing
radius---instead of rising as in \citeauthor{Dre86}'s study---although
the decline is sensibly weaker than for the gas-deficient
galaxies. These results suggest that one possible explanation for the
relationship between disk morphology and gas content (\S~\ref{type})
could be that early spirals have an orbital distribution more radially
anisotropic than late types. To test this possibility, we have inferred
the velocity dispersion profiles of the spirals subdivided into early
and late disks. Again, we find indications of radial orbits for these
two broad morphological groupings. However, the trajectories of the
galaxies in the first group does not seem to be more eccentric than
those in the second: if anything there is a hint for the opposite
effect. Not surprisingly, the kinematic behavior of the entire spiral
population is intermediate among those shown by all the previous
subdivisions. The lowest curve in the diagram, on the other hand,
displays the radial run of the velocity dispersion for the earliest
Hubble types, i.e.\ lenticulars and ellipticals, which have been
included in the cluster galaxy samples for this purpose only. These
galaxies exhibit a markedly different behavior from the spirals,
keeping an almost constant radial profile compatible with an isotropic
distribution of velocities. (Recall that the interpretations adopted
for the observed trends in the velocity dispersion curves are only
valid in the quasi-static cluster interiors.) All these findings are
yet another manifestation of the well-known fact that S and E+S0
galaxies do not share the same kinematics: late-type galaxies are
likely recent arrivals to the virialized cluster cores, which consist
essentially of ellipticals and lenticulars \citep[e.g.][]{Sod89}. In
addition to supporting this basic picture, our data also indicate that
a segregation develops among the orbits of the infalling spirals
according to their gaseous contents since the objects with the more
eccentric trajectories, \emph{regardless of morphology}, reach deeper
into the cluster cores and are thus more efficiently stripped of their
neutral hydrogen.

The same analysis for the Virgo cluster galaxies is reproduced in the
top panel of Figure~12. We see that, to a first approximation, the
velocity dispersion profiles corresponding to all the different galaxy
subgroups are essentially flat (notice, for instance, the curve
exhibited by the entire spiral population), although with a noticeably
positive excess of the velocity dispersion of the spirals relative to
the E+S0 population. As Virgo is still a dynamically young galaxy
system \citep[see e.g.][]{SBB99,Gav99}, we interpret these results as
indicative of the fact that the trajectories of the spiral galaxies
within the central Virgo region are strongly perturbed by large and
rapid fluctuations of the mean gravitational field caused by the
ongoing merger of major subclumps. Because of this large-scale phase
mixing, environmental influences on the disks have not yet been capable
of inducing a neat orbital segregation between gas-poor and gas-rich
objects. After this paper was submitted a preprint from \citet{Vol00}
became available that investigates, by means of an analytical model of
the Virgo cluster, the link between the neutral gas contents of the
cluster spirals and their orbits. Their model leads to a scenario in
which the majority of HI deficient galaxies of the Virgo centermost
region are on radial orbits and have passed through the cluster center
at least once. Interestingly enough, the presence of some high-velocity
gas-poor objects at relatively large clustercentric distances is
interpreted by these authors as a consequence of the perturbations to
the main gravitational potential arising from the radial infall of M86
towards M87. These ``special'' galaxies would result from the spirals
that populate the outskirts of the M86 cloud and that, still bound to
the system, have been scattered with high velocities to large apocenter
orbits during the merger process.

\section{Implications of the Results on the Mechanism of HI 
Depletion}\label{discussion}

The idea that a spiral galaxy moving through the hot intracluster
medium may have its HI removed by ram pressure was first introduced by
\citet{GG72} and has since been extensively invoked. The typical value
for the disk restoring force inferred in the solar neighborhood implies
that, for ram pressure to be effective, galaxies must pass through or
near the cluster cores. Under such circumstances, the low-density HI
component can be stripped fairly easily, particularly from the outer
disk regions \citep*{AMB99}. However, the molecular clouds, with
densities one million times higher and much lower filling factors,
should not be affected. This prediction is in agreement with the
observations by \citet{Sta86} and \citet{KY89} that several
HI-deficient Virgo spirals show normal molecular gas contents as
indicated by their CO luminosities. Similarly, \citet*{KBS84} found
that the distribution of H$\alpha$ equivalent widths for spirals in the
Cancer, Coma, and A1367 clusters was poorly correlated with the HI
content. Truncation of the outer HI disks is also expected from a
mechanism in which the local gravity of the galaxy plays an important
role in counteracting gas removal. Indeed, the efficiency of ram
pressure is regulated by factors such as the mass surface density of
the gas and the replenishment rate of the interstellar medium (ISM),
related to galaxy type. This complication may well explain our finding
that gas deficiency varies with morphology, while it is essentially
independent of the size of the stellar disk (\S~\ref{galprop}). In this
one respect, it is interesting to note that Sb's and earlier spirals
often exhibit central HI depressions \citep{Cay94,BvW94} which,
according to recent hydrodynamical treatments of stripping, amplify the
effectiveness of this process \citep*{MQB99}. Thus, it seems reasonable
that the strongest deficiencies correspond to the earliest spiral
disks.

Observational evidence of ongoing ISM-IGM interactions is provided by
galaxies with strongly asymmetric HI distributions
\citep[e.g.][]{DG91,Bra00}. The HI surface-density distributions of
these objects show shifts between the optical and 21-cm positions, with
extended tails combed backwards from the cluster center and a sharp
edge on the forward side, as would be expected from external dynamical
pressure effects if the galaxies were currently moving towards the
cluster center. In some cases, these asymmetries are associated with
radio continuum trails in the same direction as the HI is offset and
enhanced star formation on the compressed side of the gas disk
\citep{GJ87,DG91}. Often, the interaction of the galaxies with the
intracluster medium has been cited as one the most probable
explanations for the presence of blue galaxies with low HI contents in
the central portion of some clusters \citep[e.g.][]{BD86}. Theoretical
studies back up also the plausibility of ram pressure as the physical
mechanism behind the change in the star formation rates and colors
observed in high-redshift clusters \citep{FN99}, the morphology-density
relation of the disk galaxy population in present-day rich clusters
\citep{SS92}, and the observed HI deficiency pattern in the Virgo
cluster \citep{Vol00}.

Thermal conduction is another IGM-related mechanism capable of
producing substantial stripping rates \citep{Nul82}. Its efficiency
notwithstanding, this process has longer timescales than ram pressure
and is insensitive to the orbital parameters of the galaxies. In
addition, it depends very weakly on the galaxy's gravity, so it is
unlikely that it can generate the observed asymmetries in the HI
surface distributions. Galaxy-galaxy interactions \citep{Ick85} can
also cause important gas depletion, either directly by the tides
generated in these encounters or indirectly by inducing star
formation. The fact that the galaxy relaxation times are comparable to,
or greater than, the age of the universe has led \citet{VJ91} to
propose that tidal encounters must occur within subclumps prior to the
cluster virialization. Gravitational encounters, however, cannot remove
the HI from the inner parts of the galaxies without leaving their
imprint on the stars or the molecular component. Model calculations
show that tidal effects should produce extended tail structures both in
the stellar distribution and in the neutral hydrogen, the latter with
surface densities well above the detection threshold of the most
sensitive two-dimensional observations. This prediction is at odds with
aperture synthesis observations, which show that the HI distributions
of cluster galaxies fall off rather rapidly with respect to field
galaxies, implying a dearth of atomic gas in the outer parts. The role
of gravitational interactions on the morphology of disks---summarized
in the modern concept of ``galaxy harassment''---has been examined in
the context of hierarchical cosmogonies by
\cite{Moo96,Moo98}. According to their numerical simulations, low
surface brightness galaxies can evolve into low luminosity dwarf
spheroidals under the influence of rapid tidal encounters with giant
galaxies and cluster substructure over a timescale of several billion
years. Luminous spirals with large bulges are only affected marginally
by this process.

The natural consequence of gas losses as radical as our results
indicate would be---provided gas replenishment does not occur at
exceptionally high rates---a reduction in the star-formation activity
of the galaxy, followed by the fading of the disk. This prediction is
in good agreement with the decline of disk luminosity and the
invariance of bulge brightness with increasing local density observed
in the spirals of rich clusters \citep*{SSS89}. Also quite consistent
with this idea is the finding by \citet{KK98} that objects in Virgo
classified as Sa have similar bulge to disk ratios than the Sc's and
only differ in their overall star formation rates which are strongly
reduced in the outer disk. The morphological transformation of the
swept galaxies into S0-like objects could be completed through the
suppression of the spiral features by continued disk heating by tidal
encounters, as suggested by \citet{MQB99}. Of course, the possible
morphological evolution of cluster spirals towards earlier types has
serious difficulties in explaining the presence of S0's in the
field. While some stripped galaxies may have fairly radial orbits that
carry them at large distances from the cluster centers, one must bear
in mind that not all the lenticular galaxies, outside and inside
clusters, arise necessarily from HI-deficient spirals.

We conclude that the present investigation provides clear evidences of
the strong influence that the cluster environment has on the gaseous
disks of spirals. The marked radial pattern of HI deficiency
(\S~\ref{radial}) indicates that galaxies lose their gas near the
cluster centers. This result is consistent with the finding that
spirals with substantial HI deficiency follow orbits with large radial
components (\S~\ref{orbits}). It appears then, that the stripping of
gas requires high IGM densities and relative velocities. According to
these results, ISM-IGM interactions, basically ram pressure
supplemented by the accompanying effects of viscosity and turbulence,
are favored over other environmental interactions as the main cause of
gas depletion in clusters. The existence of very HI-deficient galaxies
in the cluster cores, often with deficiency factors of 10 or more
(\S~\ref{type}), but that look normal in other aspects (e.g.\ intrinsic
color indices, CO contents), lends weight to the conclusion that the
stripping must be relatively recent (probably a few Gyrs ago).

Qualitative support to an IGM-related stripping scenario arises also
from the HI deficiency map of the central Virgo region
(\S~\ref{vir3w}). The two main subunits of this cluster appear to be in
an advanced state of merging, so it is not surprising that their member
galaxies, which are moving through the densest portions of the gas
sitting on the cluster main potential well, exhibit the highest
deficiencies. We have detected also lumps of high HI-deficient galaxies
at large projected distances, likely related to secondary galaxy
density enhancements and/or background subclumps, which appear to be
connected with the cluster center by gas-deficient zones. We speculate
that these galaxy aggregates may have already experienced a first
high-velocity passage through the Virgo core, that could have affected
the gas content of their galaxies and left behind a trail of
gas-deficient objects, but that was insufficient to tear apart the
densest portions of the lumps. Finally, we want to point out that the
two zones having the lowest gas deficiency in our HI-deficiency
maps---a small region to the East and South of M49 and a larger one
mainly to the South of the M cloud---show a good positional
correspondence with two infalling clouds composed almost entirely
($\sim 80\%$) of spirals \citep{Gav99}.

\section{Summary}

In this paper, we have used 21-cm-line data to infer the HI contents of
1900 spiral galaxies spanning types Sa to Sdm/Irr in 18 nearby cluster
regions. Each galaxy sample is defined by a radial velocity filter of
$3\sigma$ around the systemic velocity of the central cluster and a
projected radius of 5\ra\ around the cluster center, except for the
Virgo region, in which we have included only the galaxies located
within 3\ra\ of M87.

Following our previous studies, HI deficiency has been quantified by
the difference between the observed neutral hydrogen mass and that
expected for an isolated galaxy with the same morphological type and
linear optical diameter. Improved standards of comparison are taken
from the sample of galaxies in low density environments discussed in
\citeauthor{SGH96}. The quality, sensitivity, and large size of the
dataset assembled has afforded us the possibility of making---for the
first time in a study of these characteristics---an exhaustive,
statistically rigorous investigation of the connections between gas
deficiency and the properties of both the underlying galaxies and their
environment. The main results are:

\begin{itemize}

\item Comparison of the distributions of HI content for the spiral
population in the inner ($r\le1$\ra) and outer ($r\!>\!1$\ra) portions
of each cluster field shows that twelve of the systems investigated
here---A262, A397, A400, A426, A779, A1060, A1367, Virgo, A1656, A2063,
A2147, and Pegasus---may be considered deficient in HI. Among the
non-deficient clusters, three are in the ACO catalog---A539, A3526,
and A2151---, while the other three---Pisces, N507, and Cancer---are
loosely organized galaxy concentrations.

\item The zone of HI paucity can extend out to as much as 2\ra\ from
the center of clusters. In the outskirts of these systems, the
proportion of gas-deficient objects is compatible with the field
values, while in the central cluster regions, HI deficiency is strongly
anticorrelated with the projected radial position of the galaxies.

\item The total fraction of HI-deficient spirals in a cluster shows no
statistically significant trend with other cluster global properties
such as X-ray luminosity and temperature, velocity dispersion,
richness, or spiral fraction. This result could simply be due to a
higher rate of transformation of swept spirals into lenticulars in the
richest X-ray clusters. However, possible selection effects and biases
propitiated by the small size of some of our cluster galaxy samples
make any firm conclusions about this point impossible and demand
further investigation with still larger and more complete datasets.

\item The amount of gas depletion appears to be related to the
morphology of the disks, but it is hardly a function of their optical
size. The type dependence is in the sense that both the proportion of
gas-deficient objects and the degree of depletion are higher for the
early spirals. In the Virgo cluster, where 21-cm observations expand a
large range in HI mass, most Sa's have gas deficiencies exceeding
factors of 10, a few being HI poor by more than a factor of 100. The HI
deficiency distribution for the Sdm/Irr types in this cluster shows
also a heavy tail at the high-deficiency end.

\item Orbital segregation of disks according to gas content is observed
in HI-deficient clusters: spirals devoid of gas have more eccentric
orbits than the gas-rich objects. In the dynamically young Virgo
cluster, however, no dependence of the gas deficiency on the orbital
parameters of the galaxies is discernible. Collective relaxation
effects might be responsible for the spatial pattern of HI deficiency
observed in the central Virgo region.

\end{itemize}

The progressive increase of gas deficiency towards the cluster centers,
the eccentricity of the orbits of the gas-poor galaxies, and the 2D
pattern of HI deficiency in the central Virgo region, point to a
scenario in which gas-sweeping events occur close to the cluster cores
where the density of the IGM is highest and the gas-dynamical
interactions are strongest. The detection of galaxies with extreme HI
deficiencies, but still retaining their spiral morphology, suggests
that the stripping of the atomic hydrogen is a relatively recent event
in the life of these objects. Furthermore, the details of the
relationship between HI deficiency and morphology are consistent with
the idea that the presence of central depressions in the HI disks
increases the efficiency of gas removal.

We propose that ISM-IGM dynamical interactions are the main agent
causing the ablation of the spiral disks in the cluster interiors.

\begin{acknowledgements}
We would like to thank Gregory Bothun for his careful and prompt
refereeing which has led to a much-improved presentation of the paper.
J.~M.~S. would like also to express his gratitude to Eduardo
Salvador-Sol\'e for many fruitful discussions and the Departament
d'Astronomia i Meteorologia at the Universitat de Barcelona for its
generous hospitality. J.~M.~S., A.~M., C.~G.~G., and G.~G.~C.
acknowledge support by the Direcci\'on General de Investigaci\'on
Cient\'{\i}fica y T\'ecnica, under contracts PB96--0173 and
PB97--0411. Partial support has also been provided by US NSF grants
AST--9528860 to M.~P.~H., AST--9617069 to R.~G., and AST--9900695 to
M.~P.~H. and R.~G.
\end{acknowledgements}


\clearpage
\onecolumn



\plotone{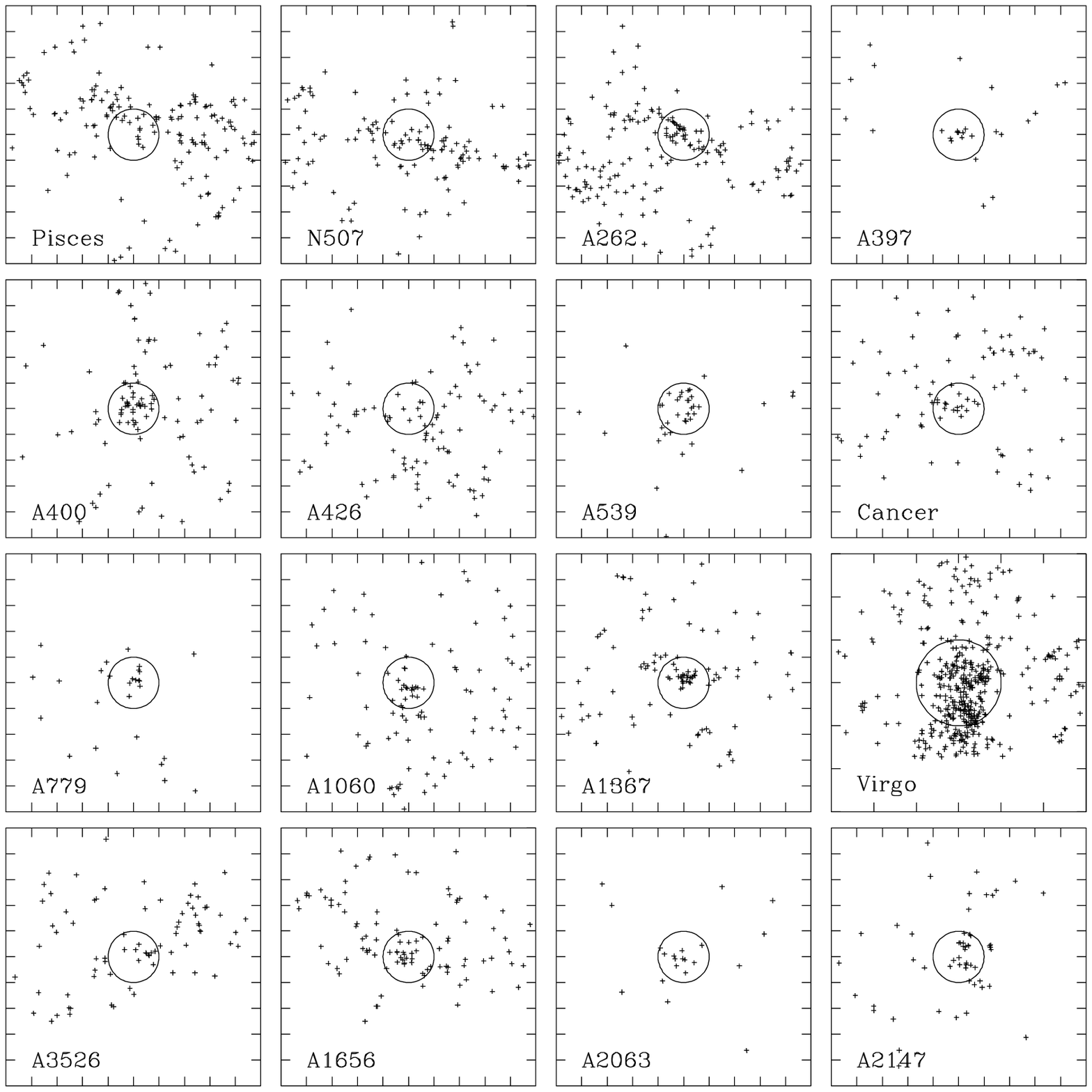} \figcaption[f1a.eps]{Sky distribution of
spirals with good HI deficiency measures in each one of our eighteen
nearby cluster fields. Tickmarks are in units of Abell radii. The
circle superposed on each panel encompasses the objects located within
the innermost 1\ra\ region. \label{f1a}}

\setcounter{figure}{0}
\plotone{f1b.eps}
\figcaption[f1b.eps]{Continued. \label{f1b}} 

\plotone{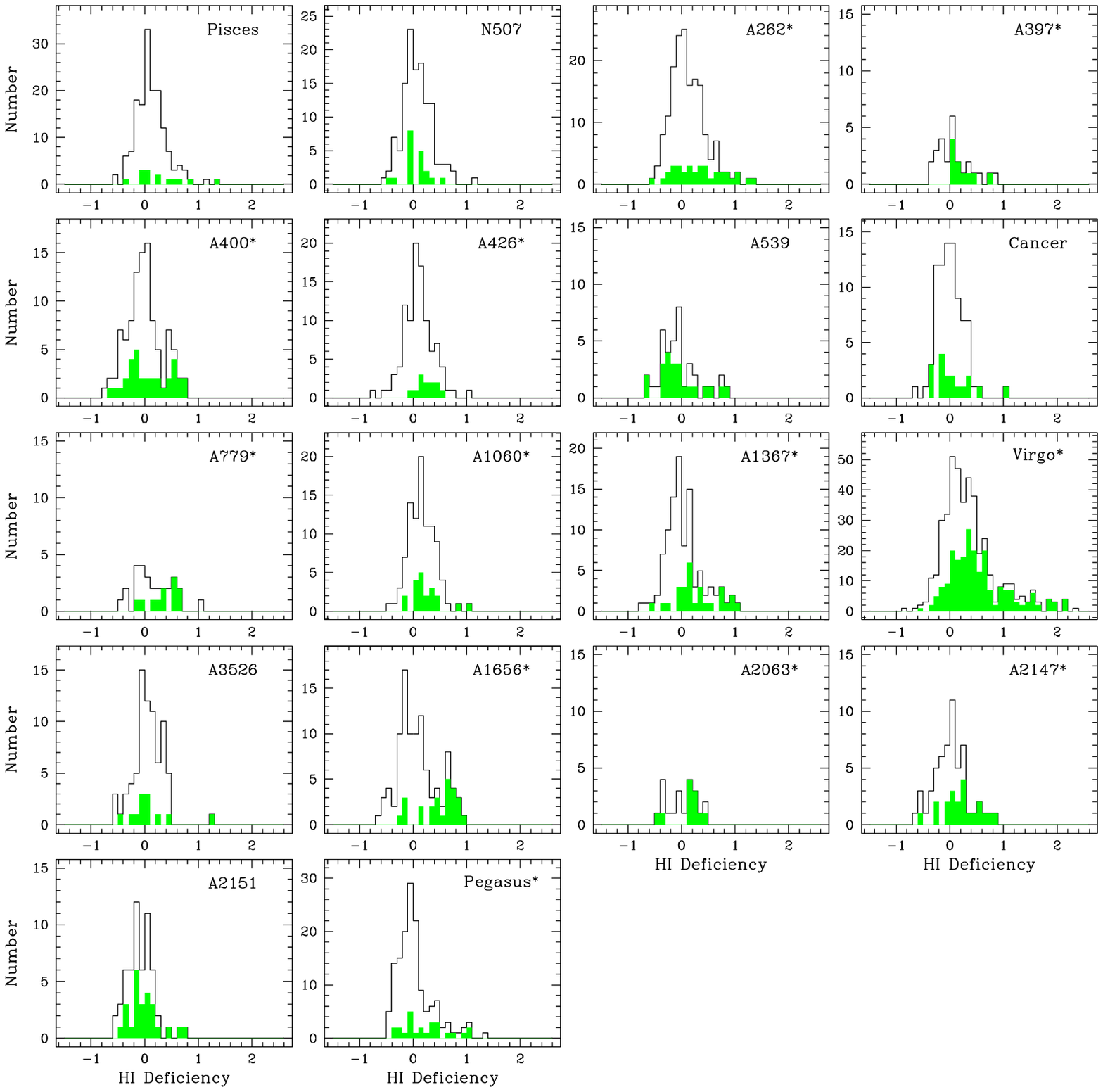} \figcaption[f2.eps]{Histograms showing the
distribution of the computed HI deficiency parameter $\df$ for each
cluster region. In each panel the filled portions of the histogram
indicate deficiencies for galaxies located within 1\ra\ of the cluster
center, while the unfilled areas represent galaxies at larger radii: up
to 3\ra\ for the Virgo field and up to 5\ra\ for the remaining cluster
regions. HI-deficient clusters (identified by an asterisk after the
name) are those for which a Kolmogorov-Smirnov test finds less than
$10\%$ probability that the inner and outer distributions of $\df$ are
drawn from the same parent population. \label{f2}}

\plotone{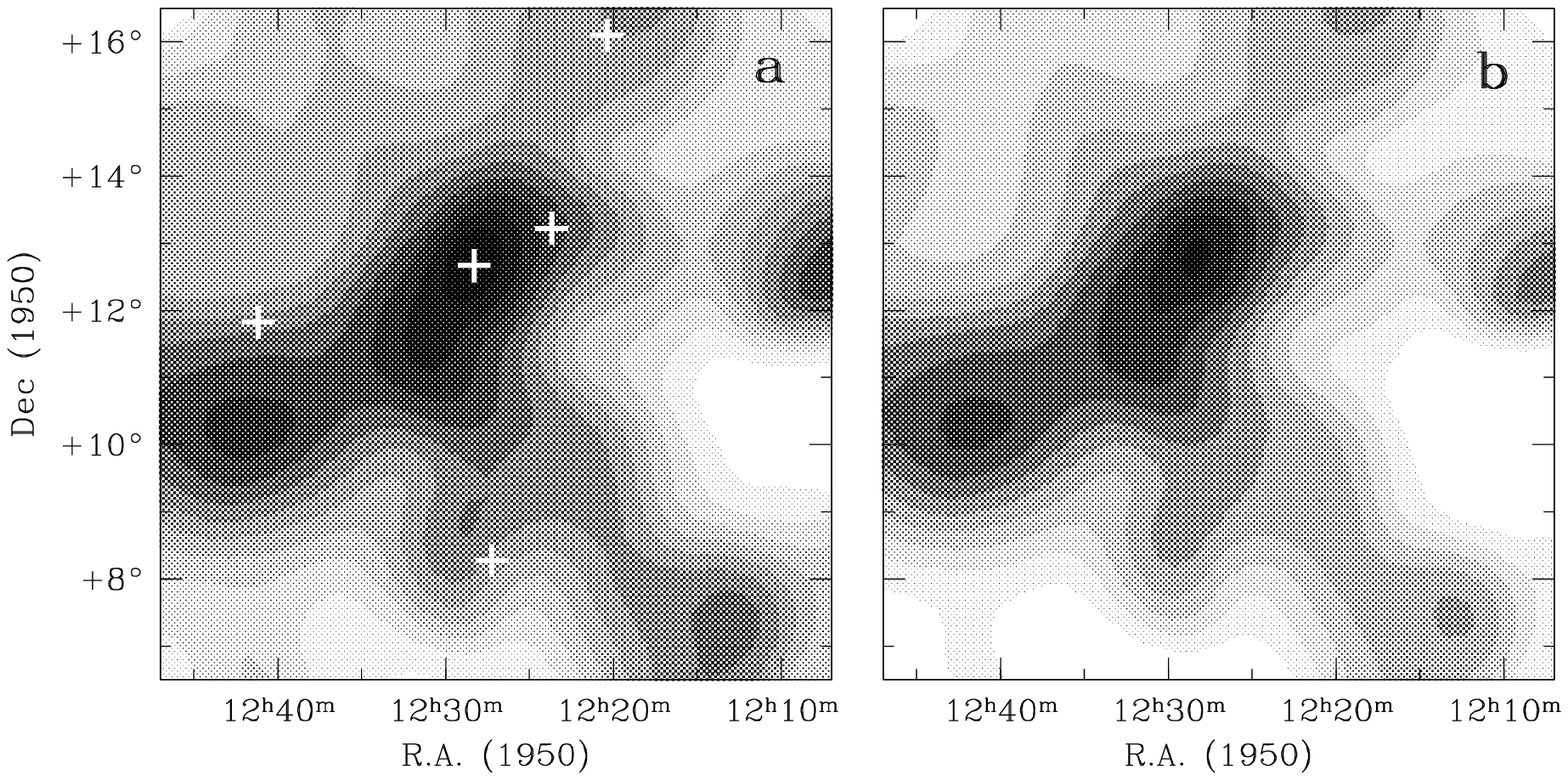} \figcaption[f3.eps]{Greyscale images of the sky
distribution of the HI deficiency in the central region of the Virgo
cluster: (a) using the HI deficiency parameter $\df$, and (b) a
distance-independent approximation of it (see text). The lightest
regions in the maps correspond to $\df<0.1$ and the darkest to $\df\ge
1.2$. The contour spacing is linear. The peak value of HI deficiency is
located near the position of M87, which coincides with the maximum of
X-ray emission in the whole area. The positions of five dominant
galaxies are marked by crosses (top to bottom: M100, M86, M87, M60, and
M49). The size of the images is $10\degr\times 10\degr$. \label{f3}}

\plotone{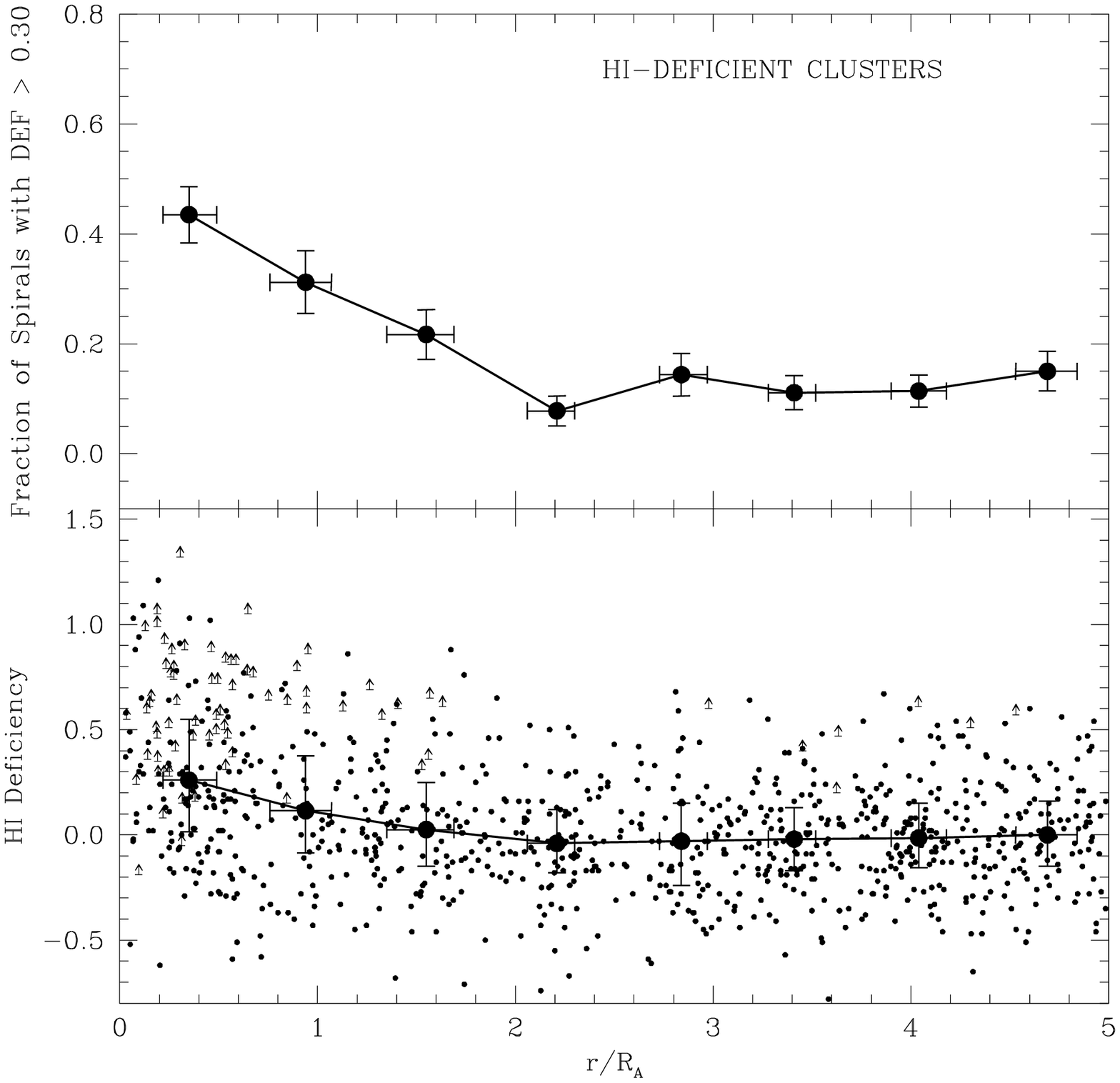} \figcaption[f4.eps]{\emph{Top:} HI-deficient fraction
in bins of projected radius from the cluster center for the
superposition of all the HI-deficient clusters but Virgo. Vertical
error bars correspond to $1\sigma$ confidence Poisson intervals. The
abscissas show medians and quartile values of the bins in radial
distance. \emph{Bottom:} same in upper panel for the measured HI
deficiency. Displayed are the medians and quartiles of the binned
number distributions in HI deficiency. Small dots show the radial
variation of HI deficiency for individual galaxies while the arrows
identify non-detections plotted at their estimated lower
limits. \label{f4}}

\plotone{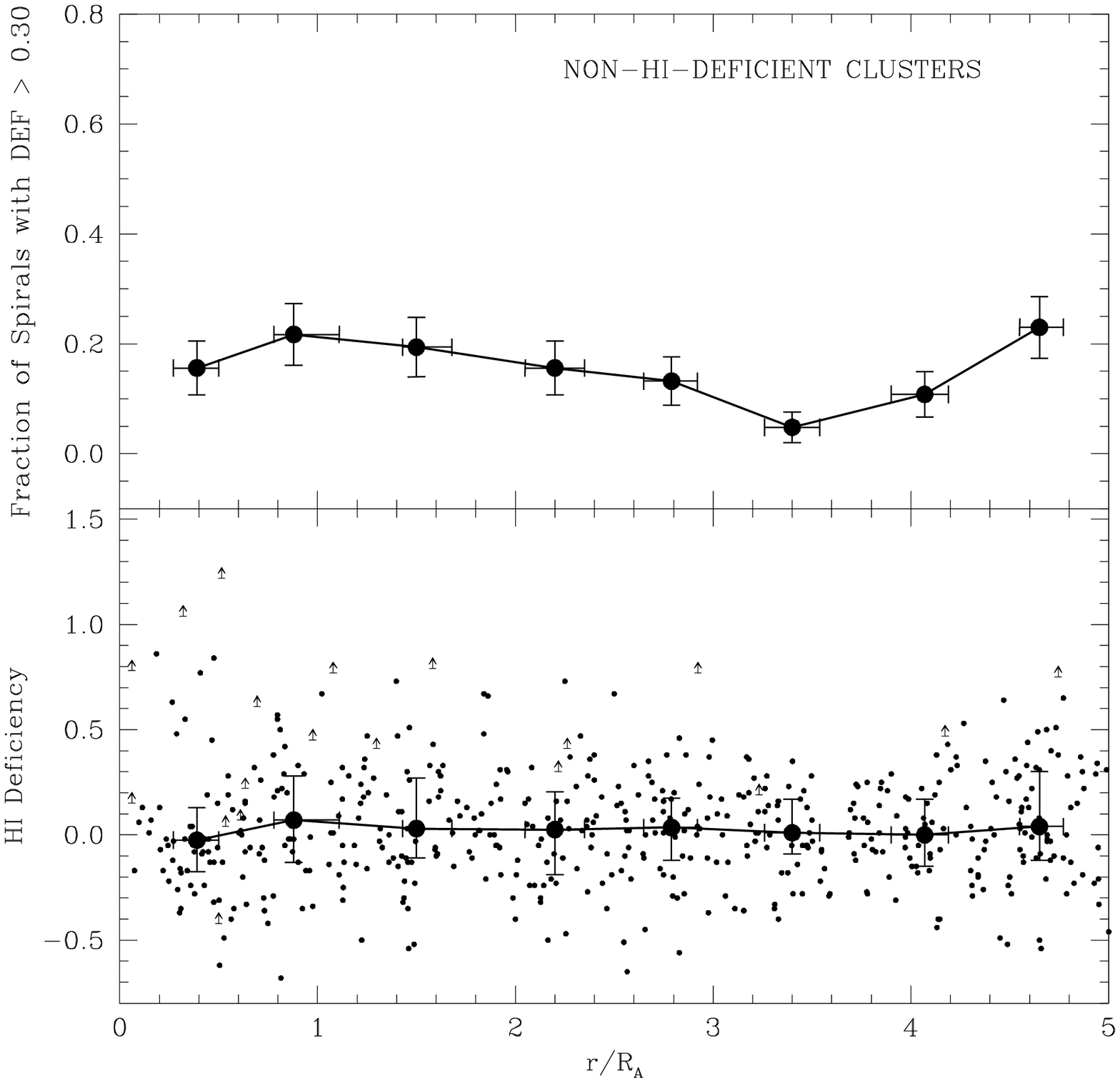} \figcaption[f5.eps]{Same as in Fig.~4 but for the
superposition of all the non-HI-deficient clusters. \label{f5}}

\plotone{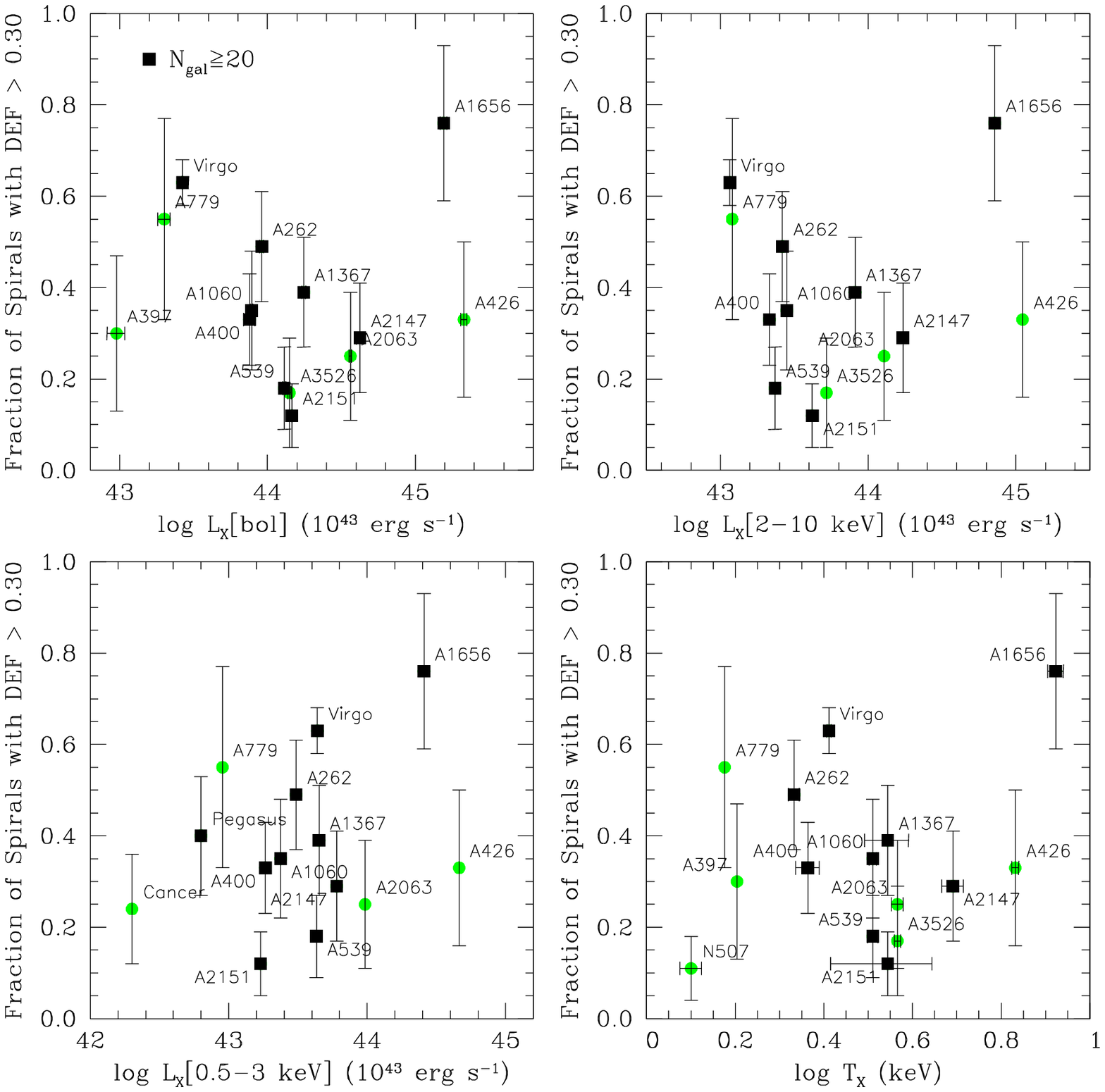} \figcaption[f6.eps]{From left to right and top to
bottom, spiral fraction within 1\ra\ with a deficiency parameter $\df$
larger than 0.30 vs.\ cluster bolometric, 2--10~keV, and 0.5--3.0~keV
X-ray luminosities, and cluster X-ray temperature. Square symbols
identify clusters with a minimum of 20 objects in the central
region. Vertical error bars correspond to $1\sigma$ Poisson confidence
intervals, except for the temperature where the quoted uncertainties
are 90\% for the ASCA observations and 68\% for the \emph{Einstein}
estimates (see Table~2). \label{f6}}

\plotone{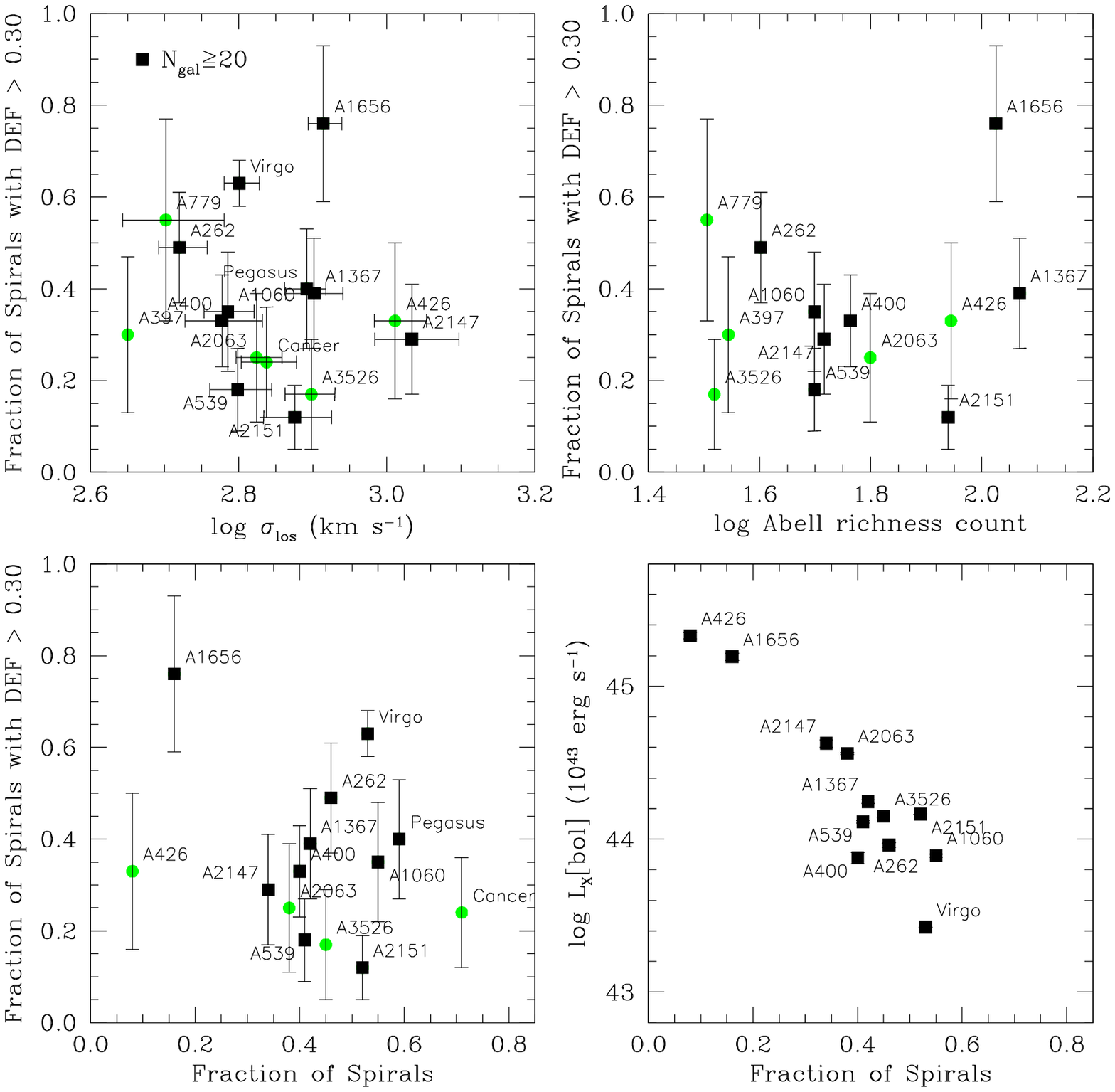} \figcaption[f7.eps]{From left to right and top to
bottom, spiral fraction within 1\ra\ with a deficiency parameter $\df$
larger than 0.30 vs.\ cluster velocity dispersion, Abell richness
count, and total fraction of spirals. The bottom right panel shows the
bolometric X-ray luminosity plotted against the total fraction of
spirals. Vertical error bars correspond to $1\sigma$ confidence
intervals. In the bottom right panel the size of the symbols is larger
than the error bars. \label{f7}}

\plotone{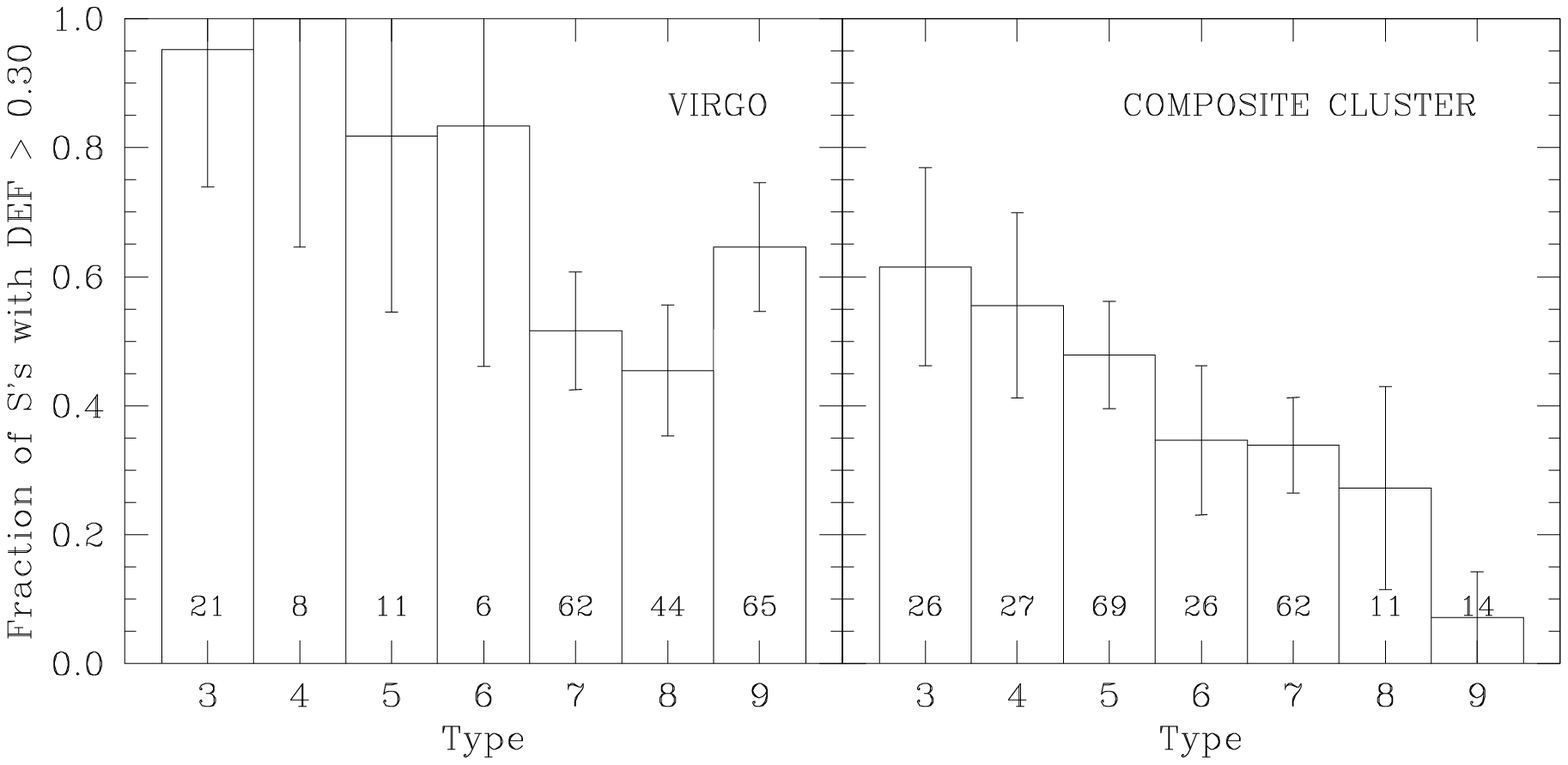} \figcaption[f8.eps]{Fraction of galaxies within
1\ra\ with deficiency parameter $\df>0.30$ as a function of the
morphological type (Sa--Sdm/Irr) for the Virgo sample (left) and a 
composite cluster sample formed by combining the rest of the HI-deficient
systems (right). The numbers within each bin indicate the total number
of galaxies in that bin. Error bars correspond to $1\sigma$ Poisson
errors.\label{f8}}

\plotone{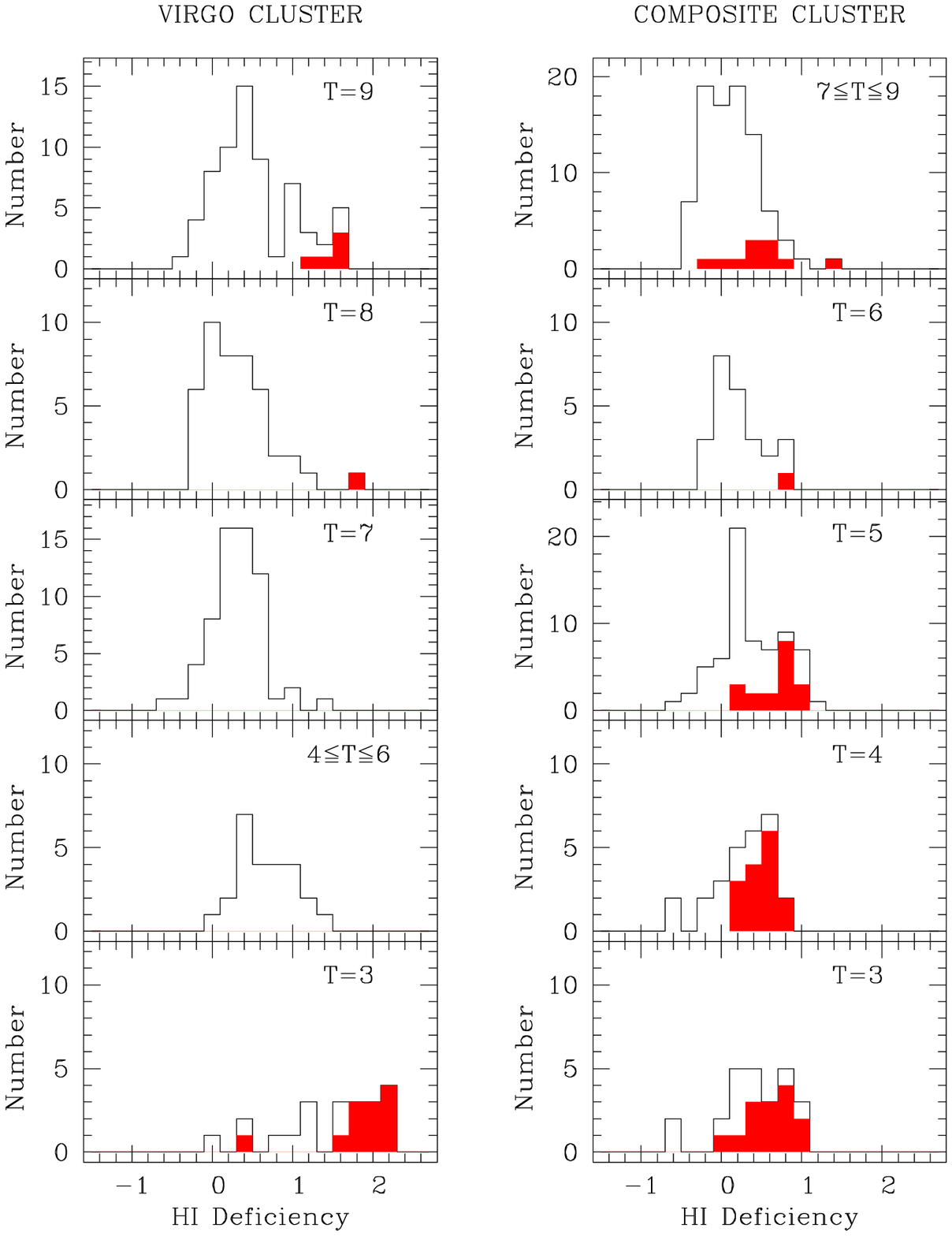} \figcaption[f9.eps]{HI deficiency histograms of the
spirals lying within 1\ra\ of the Virgo (\emph{left}) and the composite
HI-deficient cluster (\emph{right}) centers separately by morphological
type. In each panel the dark portions of the histograms correspond
to those galaxies undetected in HI for which a lower limit on HI
deficiency has been estimated. Non-detections contribute with their
nominal lower limits of deficiency.\label{f9}}

\plotone{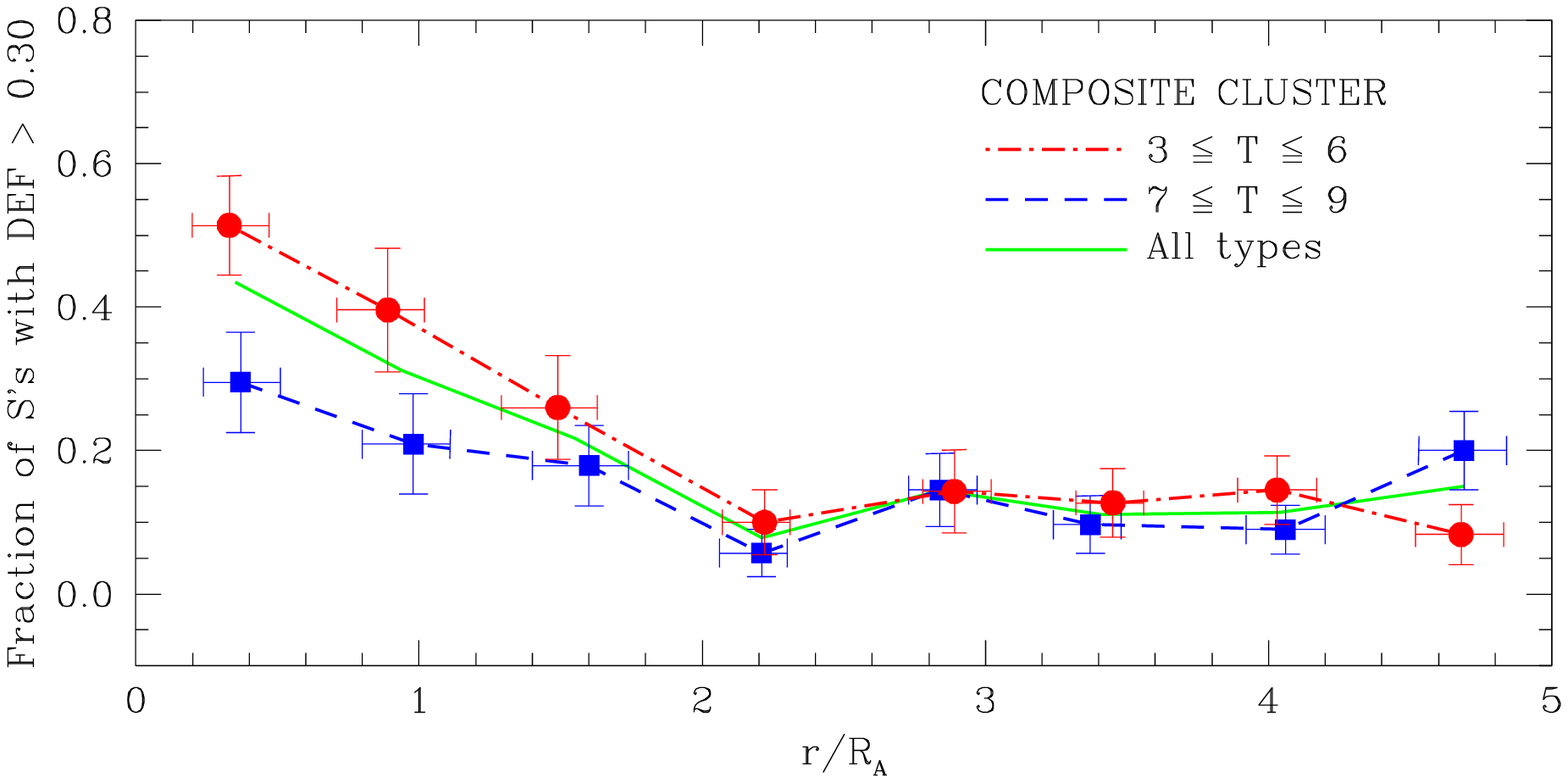} \figcaption[f9.eps]{Same as in the upper panel of
Fig.~4, for the early (circles) and late (squares) spirals
separately. The solid curve reproduces the trend of the entire spiral
population. Only error bars for the two morphological subgroups are
displayed for clarity.\label{f10}}

\plotone{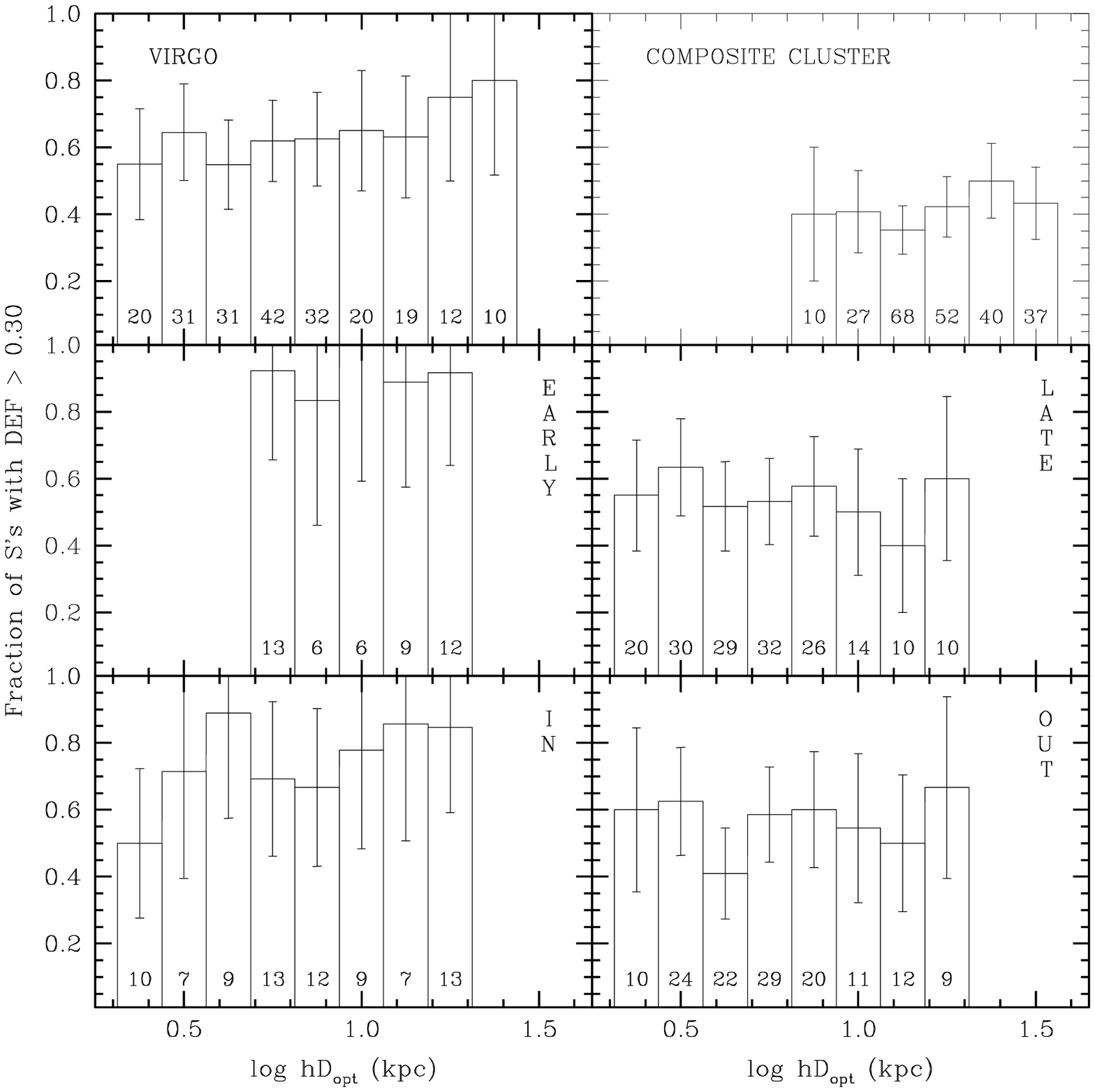} \figcaption[f11.eps]{\emph{Top:} Spiral fraction
within 1\ra\ with $\df>0.30$ as a function of the linear optical
diameter for the Virgo (left) and the composite HI-deficient cluster
(right). \emph{Middle:} Same as in the upper left panel but separately
for the early (left) and late (right) spiral subsets. \emph{Bottom:}
Same as in the upper left panel but separately for the subsets of
galaxies located within 0.5\ra\ from M87 (left) and beyond this
distance (right). The numbers within the bars indicate the galaxies in
each bin. In all panels the few galaxies with sizes outside the plotted
range have been accumulated in the extremal bins to avoid large
statistical fluctuations. Error bars correspond to $1\sigma$ Poisson
errors.\label{f11}}

\plotone{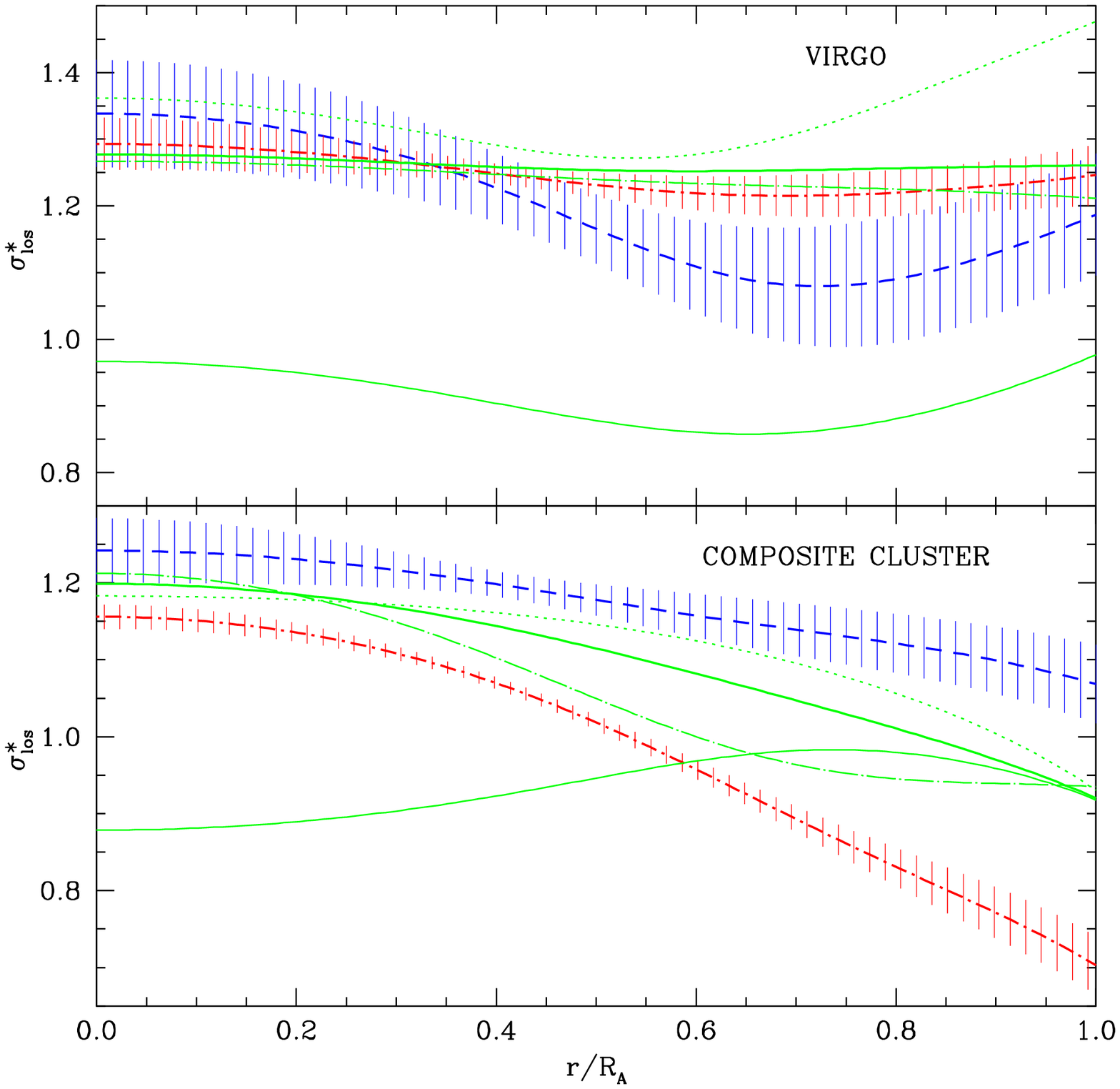} \figcaption[f12.eps] {Radial run of the
\emph{normalized} los velocity dispersion up to 1\ra\ for the Virgo
(top) and the composite HI-deficient cluster (bottom). Line coding is
as follows: thick dot-dash for spirals with $\df\ge 0.48$, thick dash
for spirals with $\df\le 0$, thick solid for all spirals, dots for
early spirals, dot-long dash for late spirals, and solid for
ellipticals and lenticulars. In both plots vertical error bars
correspond to $1\sigma$ confidence intervals. Only error bars for the
profiles corresponding to the spirals with extremal HI contents are
displayed for clarity.\label{f12}}




\clearpage 

\input{tab1}

\clearpage
\begin{deluxetable}{lrcccrcrc}
\small
\tablewidth{0pt}
\tablenum{2}
\tablecaption{Global Properties of Clusters \label{tab2}}
\tablecolumns{9}
\tablehead{\colhead{} & \multicolumn{1}{c}{$L_{\mathrm X}$(bol)\tablenotemark{a}}&
\multicolumn{1}{c}{$L_{\mathrm X}$(soft)\tablenotemark{b}} & 
\multicolumn{1}{c}{$L_{\mathrm X}$(hard)\tablenotemark{c}} & 
\colhead{$T_{\mathrm X}$\tablenotemark{d}} & 
\colhead{$\sigma_\mathrm{los}$\tablenotemark{e}} &
\colhead{} &\colhead{} &\colhead{} \\ 
\colhead{Name} &  &\multicolumn{1}{c}{(\lxu)} 
   & & 
\colhead{(keV)} & \colhead{(\kms)} &
\colhead{C\,\tablenotemark{f}} & \colhead{$F_{\mathrm S}$\tablenotemark{g}}& \colhead{\fdef}\\
\multicolumn{1}{c}{(1)} & \multicolumn{1}{c}{(2)} &
\multicolumn{1}{c}{(3)} & \multicolumn{1}{c}{(4)} & 
\multicolumn{1}{c}{(5)} & \multicolumn{1}{c}{(6)} & 
\multicolumn{1}{c}{(7)} & \multicolumn{1}{c}{(8)} &
\multicolumn{1}{c}{(9)}
}
\startdata
 Pisces&\nodata\phm{.00}&\phn\nodata&\phd\nodata&\nodata&\nodata\phd&\nodata&\nodata&$0.36\pm 0.16$\\
 N507&\nodata\phm{.00}&\phn\nodata&\phd\nodata&$1.26\pm 0.07$&\nodata\phd&\nodata&\nodata&$0.11\pm 0.07$\\
 A262&$9.15 \pm 0.30$ &\phm{$<$}3.07&\phn 2.63&$2.15\pm 0.06$&$525^{+47}_{-33}$ &40&0.46&$0.49\pm 0.12$\\
 A397&$0.95 \pm 0.13$ &\phn\nodata&\phd\nodata&1.60&447&35 &\nodata&$0.30\pm 0.17$\\
 A400&$7.56 \pm 0.65$ &\phm{$<$}1.84 &\phn 2.14&$2.31\pm 0.14$ &$599^{+80}_{-65}$ & 58 &0.40 &$0.33\pm 0.10$\\
 A426&$214.0\pm 10.3$ &\phm{$<$}46.1 &\ \ \ 111&$6.79\pm 0.12$ &$1026^{+106}_{-64}$ & 88 &0.08 &$0.33\pm 0.17$\\
 A539&$13.00\pm 0.28$ &$<$4.30 &\phn 2.34&$3.24\pm 0.09$ &$629^{+70}_{-52}$ &  50 &0.41 &$0.18\pm 0.09$\\
 Cancer&\nodata\phm{.00}&$<$0.20 &\phd\nodata&\nodata&$688^{+67}_{-52}$&\nodata&0.71 &$0.24\pm 0.12$\\
 A779&$2.00 \pm 0.19$ &$<$0.90 &$<$1.20 \ &1.50&$503^{+100}_{-63}$& 32 &\nodata&$0.55\pm 0.22$\\
 A1060&$7.81 \pm 0.18$ &\phm{$<$}2.37 &\phn 2.81&$3.24\pm 0.06$ &$610^{+52}_{-43}$ &  50 &0.55 &$0.35\pm 0.13$\\
 A1367&$17.60\pm 0.52$ &\phm{$<$}4.49 &\phn 8.15&$3.50\pm 0.40$ &$798^{+75}_{-68}$ &117\phn&0.42 &$0.39\pm 0.12$\\
 Virgo&2.66\phm{.000}&\phm{$<$}4.37 &\phn 1.16&$2.58\pm 0.03$ &$632^{+41}_{-29}$&\nodata&0.53 &$0.63\pm 0.05$\\
 A3526&$14.10\pm 0.32$ &\phn\nodata&\phn 5.21&$3.68\pm 0.06$ &$791^{+60}_{-62}$ &33 &0.45 &$0.17\pm 0.12$\\
 A1656&$156.0\pm 8.30$ &\phm{$<$}25.7&\phn 71.9&$8.38\pm 0.34$ &$821^{+49}_{-38}$ &106\phn&0.16 &$0.76\pm 0.17$\\
 A2063&$36.40\pm 0.66$ &\phm{$<$}9.65 &\phn 12.8&$3.68\pm 0.11$ &$667^{+55}_{-41}$ &63 &0.38 &$0.25\pm 0.14$\\
 A2147&$42.30\pm 2.32$ &\phm{$<$}6.03 &\phn 17.2&$4.91\pm 0.28$ &$1081^{+170}_{-117}$&52 &0.34 &$0.29\pm 0.12$\\
 A2151&14.6\phm{.000}&\phm{$<$}1.70 &\phn 4.18&$3.50\pm 0.90$ &$751^{+91}_{-69}$ &  87 &0.52 &$0.12\pm 0.07$\\
Pegasus&\nodata\phm{.00}&$<$0.63&\phn\nodata&\nodata&$780^{+47}_{-52}$&\nodata&0.59 &$0.40\pm 0.13$\\
\enddata
\tablenotetext{a}{Bolometric X-ray luminosities primarily from \citet*{WJF97}. The value for Virgo comes from \citet{Dav93}.}
\tablenotetext{b}{\emph{Einstein} X-ray luminosities in the 0.5$-$3 keV energy range from \citet{JF84} and \citeauthor{GH85}. Upper limits for A539 and A779 are from \citet{AK83}.}
\tablenotetext{c}{X-ray luminosities in the 2$-$10 keV energy range primarily from the compilation of \emph{Einstein}, \emph{Exosat}, and \emph{Ginga} data by \citet{Dav93}. The upper limit for A779 comes from \citet{QM82}.}
\tablenotetext{d}{X-ray temperatures primarily from the compilation of ASCA observations by \citet*{HMS99}. For A397, A779, A1367, and A2151, we use \emph{Einstein} estimates from \citet{WJF97}.}
\tablenotetext{e}{Cluster radial velocity dispersions primarily from \citet{Fad96}. Values for A779 and A2147 have been drawn from \citet{Zab93}, and for A397 from \citet{SR91}. Cancer was observed by \citet{Bot83}.}  
\tablenotetext{f}{Abell richness count inside 1\ra\ from \citet{ACO89}.}
\tablenotetext{g}{Fraction of spirals from averages of the values listed in \citet{Bah77}, \citet{Dre80}, and \citeauthor{GH85}.}
\end{deluxetable}

\end{document}

%% file: tab1.tex



\begin{deluxetable}{lrclcccrcccr}
\small
\tablewidth{0pt}
\tablenum{1}
\tablecaption{List of Cluster Fields \label{tab1}}
\tablecolumns{12}
\tablehead{\multicolumn{1}{c}{} & \multicolumn{3}{c}{R.A. (1950)} &
\multicolumn{3}{c}{Dec. (1950)} & \multicolumn{1}{c}{Velocity Filter} & 
\multicolumn{1}{c}{\ra} & \multicolumn{2}{c}{\# of Galaxies} & 
\multicolumn{1}{c}{}\\ 
\multicolumn{1}{c}{Name} & \multicolumn{1}{c}{(h)} &
\multicolumn{1}{c}{(m)} & \multicolumn{1}{c}{(s)} & 
\multicolumn{1}{c}{(\degr)} & \multicolumn{1}{c}{(\arcmin)} & 
\multicolumn{1}{c}{(\arcsec)} & \multicolumn{1}{c}{(\kms)} & 
\multicolumn{1}{c}{($\deg$)} & \multicolumn{1}{c}{$r\le 1R_{\mathrm A}$} & 
\multicolumn{1}{l}{$r\le 5R_{\mathrm A}\!$\tablenotemark{a}} & 
\multicolumn{1}{c}{$P_{\rm KS}$}\\
\multicolumn{1}{c}{(1)} & \multicolumn{3}{c}{(2)\phn} &
\multicolumn{3}{c}{\phn(3)} & \multicolumn{1}{c}{(4)} & 
\multicolumn{1}{c}{(5)} & \multicolumn{1}{c}{(6)} & 
\multicolumn{1}{c}{(7)} & \multicolumn{1}{c}{(8)}
}
\startdata
Pisces &00&59&54.0& $+$30 & 02 & 00 & 3500$\div$7500$\ $ & 1.88 & $\ $14 & 155 & 0.316\\ 
N507 &01&24&26.8& $+$34 & 03 & 35 & 3500$\div$7500$\ $ & 1.83 & $\ $19 & 124 & 0.497\\ 
A262 &01&49&49.9& $+$35 & 53 & 50 & 3000$\div$7000$\ $ & 1.90 & $\ $35 & 168 & 0.002\\ 
A397 &02&53&51.2& $+$15 & 41 & 35 & 8500$\div$11000$\ $ & 0.95 & $\ $10 & $\ $26 & 0.003\\
A400 &02&55&00.0& $+$05 & 48 & 25 & 5000$\div$9000$\ $ & 1.28 & $\ $33 & 100 & 0.057\\ 
A426 &03&16&30.0& $+$41 & 20 & 00 & 2000$\div$9000$\ $ & 1.70 & $\ $12 & $\ $99 & 0.008\\ 
A539 &05&13&55.2& $+$06 & 23 & 16 & 6500$\div$10500$\ $ & 1.03 & $\ $22 & $\ $38 & 0.909\\
Cancer &08&17&00.0& $+$21 & 11 & 00 & 2500$\div$7000$\ $ & 1.75 & $\ $17 & $\ $83 & 0.581\\
A779 &09&16&44.3& $+$33 & 57 & 18 & 4500$\div$9000$\ $ & 1.25 & $\ $11 & $\ $28 & 0.012\\ 
A1060 &10&34&27.7& $-$27 & 16 & 26 & 2000$\div$5500$\ $ & 2.16 & $\ $20 & $\ $96 & 0.060\\
A1367 &11&42&04.6& $+$19 & 59 & 14 & 4000$\div$9000$\ $ & 1.32 & $\ $28 & 100 & $<$0.001\\
Virgo &12&28&18.0& $+$12 & 40 & 00 & $-$500$\div$2700$\ $ & 7.23 & 218 & 426 & $<$0.001\\
A3526 &12&46&06.0& $-$41 & 02 & 00 & 1400$\div$4500$\ $ & 2.54 & $\ $12 & $\ $76 & 0.491\\
A1656 &12&57&18.3& $+$28 & 12 & 22 & 4000$\div$10000$\ $ & 1.25 & $\ $25 & 100 & $<$0.001\\ 
A2063 &15&20&39.1& $+$08 & 47 & 18 & 9000$\div$12000$\ $ & 0.84 & $\ $11 & $\ $21 & 0.004\\
A2147 &15&59&58.3& $+$16 & 06 & 15 & 8000$\div$14000$\ $ & 0.86 & $\ $21 & $\ $57 & 0.002\\
A2151 &16&02&22.0& $+$17 & 51 & 48 & 8000$\div$14000$\ $ & 0.82 & $\ $25 & $\ $58 & 0.405\\
Pegasus &23&18&00.0& $+$07 & 55 & 00 & 2000$\div$5500$\ $ & 2.36 & $\ $25 & 145 & 0.011\\ 
\enddata
\tablenotetext{a}{The maximum radial cutoff for the Virgo cluster is 3\ra.}
\end{deluxetable}